*To be submitted to Physical Review Applied*

# Effect of surface ionic screening on polarization reversal and phase diagrams in thin antiferroelectric films for information and energy storage

Anna N. Morozovska[1*], Eugene A. Eliseev[2], Arpan Biswas[3], Nicholas V. Morozovsky[1], and Sergei V. Kalinin[3†]

[1] Institute of Physics, National Academy of Sciences of Ukraine, pr. Nauky 46, 03028 Kyiv, Ukraine

[2] Institute for Problems of Materials Science, National Academy of Sciences of Ukraine, Krjijanovskogo 3, 03142 Kyiv, Ukraine

[3] Center for Nanophase Materials Sciences, Oak Ridge National Laboratory, Oak Ridge, TN 37831

**Abstract**

The emergent behaviors in antiferroelectric thin films due to a coupling between surface electrochemistry and intrinsic polar instabilities are explored within the framework of the modified 2-4-6 Kittel-Landau-Ginzburg-Devonshire (KLGD) thermodynamic approach. Using phenomenological parameters of the KLGD potential for a bulk antiferroelectric ($PbZrO_3$) and a Stephenson-Highland (SH) approach, we study the role of surface ions with a charge density proportional to the relative partial oxygen pressure on the dipole states and their reversal mechanisms in the antiferroelectric thin films. The combined KLGD-SH approach allows to delineate the boundaries of antiferroelectric, ferroelectric-like antiferroionic and electret-like paraelectric states as a function of temperature, oxygen pressure, surface ions formation energy and concentration, and film thickness. This approach also allows the characterization of the polar and antipolar orderings dependence on the voltage applied to the antiferroelectric film, and the analysis of their static and dynamic hysteresis loops. The applications of the antiferroelectric films covered with surface ion layer for energy and information storage are explored.

[*] corresponding author, e-mail: anna.n.morozovska@gmail.com
[†] corresponding author, e-mail: sergei2@ornl.gov

# I. INTRODUCTION

Ferroelectric (**FE**) phase stability requires effective screening of the polarization bound charge at surfaces and interfaces with non-zero normal component of polarization [1, 2, 3]. Rapid growth of FE thin film applications in 90ies necessitated the analysis of the microscopic mechanisms acting at ferroelectric interfaces, preponderantly effects stemming from non-zero spatial separation between a spontaneous polarization and screening charges [4, 5, 6, 7, 8]. These effects are often introduced via the dead layer [2] or physical gap [9] concepts, postulating the presence of the thin non-ferroelectric layer or gap separating the ferroelectric surface from the electrode. The validity of this approximation was confirmed by the microscopic density functional theory studies [10, 11, 12].

However, the dead layer approximation largely ignores the realistic details of the screening process at the open surfaces and ferroelectric-semiconductor interfaces. In particular, the stabilization of FE state in ultrathin perovskite films can take place due to the chemical screening (see e.g. [13, 14, 15]), and the screening via ionic adsorption is intrinsically coupled to the surface electrochemical processes [16, 17, 18]. This screening mechanisms was confirmed by multiple experimental observations including the polarization retention above Curie temperature [19], temperature induced domain potential inversion [20], formation of bubble domains during tip-induced switching [21, 22, 23], chaotic switching [7] and domain shape instabilities [24]. Macroscopically, it is confirmed via the chemical switching in ferroelectrics [13, 25]. Finally, multiple anomalous observations such as tip pressure induced switching [26, 27] or continuous polarization states in ultrathin films can be partially attributed to the ionic screening [28]. This coupling results in non-trivial influence on the FE phase stability and phase diagrams [25, 29], albeit the overall research effort in this area is fairly small.

The early theoretical analyses, though studied the properties of ferroelectric material in details, typically ignored the nonlinear tunable characteristics of surface screening charges. A complementary thermodynamic approach was developed by Stephenson and Highland (**SH**) [25, 29], who consider an ultrathin film in equilibrium with a chemical environment that supplies (positive and negative) ionic species to compensate its polarization bound charge at the surface.

Recently, we modified the SH approach allowing for the presence of a gap between the ferroelectric surface covered by ions and a SPM tip [30, 31, 32, 33, 34], and developed the analytical description for thermodynamics and kinetics of these systems. The analysis [30 – 34] leads to the elucidation of ferroionic states, which are the result of nonlinear electrostatic interaction between the ions with the surface charge density obeyed Langmuir adsorption isotherm and ferroelectric dipoles. The properties of these states were described by the system of coupled 1D equations and corresponding phase diagrams have been established.

Here we study the role of the surface ions with a charge density proportional to the partial oxygen pressure on the dipole states and its reversal mechanisms, and corresponding phase diagrams of antiferroelectric (**AFE**) thin films. We use the SH approach combined with the 2-4-6 Kittel-Landau-Ginzburg-Devonshire (**KLGD**) thermodynamic potential for two polarization sublattices for the description of the polar and antipolar long-range orderings. Appeared, that, compared to FE materials, the considered AFE systems reveal more complex dynamics of polarization and surface charge due to presence of the FE-like antiferroionic (**AFI**) states. We analyze corresponding phase diagrams and associated hysteresis loops, as well as the application of the AFE films covered by surface ion layer for information and energy storage, and multi-bit nonvolatile random-access memory.

The manuscript is structured as following. **Section II** contains basic KLGD equations and SH problem formulation with boundary conditions. **Section III** contains free energy of the considered system. Numerical results (phase diagrams, hysteresis loops), their discussion and analysis are presented in **Section IV**. **Section V** is an outlook and brief summary. Parameters used in calculations and auxiliary figures are listed in **Appendices A-E** of **Suppl. Mat**. [35].

## II BASIC EQUATIONS WITH BOUNDARY CONDITIONS

Here we consider the system consisting of an electron-conducting bottom electrode, an AFE film, and a layer of surface ions with a charge density $\sigma(\phi)$. An ultra-thin gap separates the film surface and the top electrode, that is either ion-conductive planar electrode or flatted apex of SPM tip. The electrode provides a direct ion exchange with an ambient media, as shown in **Fig. 1(a).** A mathematical statement of the problem is listed in **Appendix A** [35]**.**

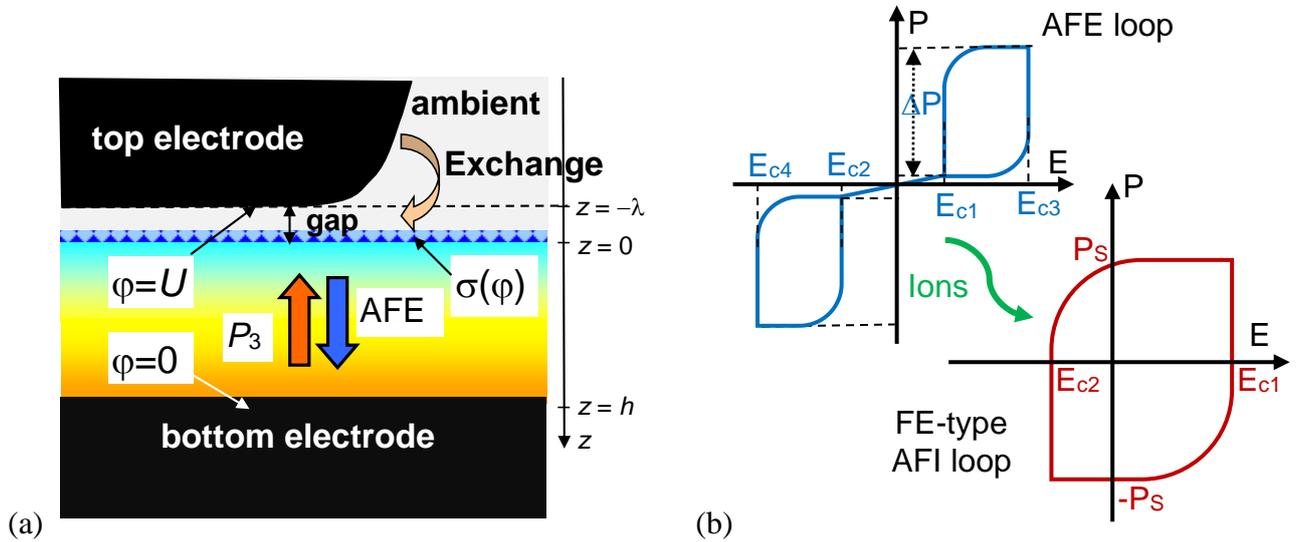

**FIGURE 1**. **(a)** Layout of the considered system, consisting of an electron-conducting bottom electrode, an AFE film of thickness $h$, a layer of surface ions with a charge density $\sigma(\phi)$, an ultra-thin gap separating film surface, and a top electrode providing a direct ion exchange with an ambient media (from bottom to the top). The film thickness is $h$, the gap thickness is λ. Adapted from Ref. [31]. **(b)** Schematics of the transition from an AFE-type polarization hysteresis to an FE-type AFI hysteresis loop induced by electric coupling with the charge of surface ions.

Due to the presence of an ultra-thin dielectric gap between the top electrode and the surface of AFE film, the linear equation of state $\boldsymbol{D} = \varepsilon_0\varepsilon_d \boldsymbol{E}$ relates an electric displacement $\boldsymbol{D}$ and field $\boldsymbol{E}$ in the gap. Here $\varepsilon_0$ is a universal dielectric constant and $\varepsilon_d \sim (1 - 10)$ is a relative permittivity in the gap filled by an air with controllable oxygen pressure. A wide band-gap AFE film can be considered insulating, and here $\boldsymbol{D} = \varepsilon_0 \boldsymbol{E} + \boldsymbol{P}$. A potential $\phi$ of a quasi-static electric field inside the film satisfies a Laplace equation in the gap and a Poisson equation in the film. The boundary conditions for the system are the equivalence of the electric potential to the applied voltage $U$ at the top electrode (modeled by a flattened region $z = -\lambda$); and the equivalence of the difference $D_3^{(gap)} - D_3^{(film)}$ to the ionic surface charge density $\sigma[\phi(\vec{r})]$ at $z = 0$; the continuity of the $\phi$ at gap - film interface $z = 0$; and zero potential at the conducting bottom electrode $z = h$ [see **Fig. 1**].

The polarization components of the uniaxial AFE film depend on the electric field $E_i$ as $P_i = \varepsilon_0(\varepsilon_{ii}^f - 1)E_i$ and $P_3 = P_3^f + \varepsilon_0(\varepsilon_{33}^b - 1)E_3$, where $\varepsilon_{ii}^f$ is a relative dielectric permittivity of AFE, $i = 1,2$, and $\varepsilon_{33}^b$ is a relative background permittivity of antiferroelectric, $\varepsilon_{33}^b \leq 10$ [36].

The polarization component $P_3^f$ is further abbreviated as $P_3$. Below we use the phenomenological two-sublattice Kittel model [37] as the basic framework for description of AFE material. Within the model we introduce the polar and antipolar long-range order parameters, $P = \frac{1}{2}\left(P_3^{(1)} + P_3^{(2)}\right)$ and $A_3 = \frac{1}{2}\left(P_3^{(1)} - P_3^{(2)}\right)$, where $P_3^{(j)}$ is the normal component of the $j$-th sublattice polarization. We combine the Kittel model with a 2-4-6-Landau-Ginzburg-Devonshire thermodynamic potential for the description of long-range polar and antipolar orderings. Hence, the evolution and spatial distribution of the polar ($P$) and antipolar ($A$) order parameters are given by the coupled time-dependent 2-4-6 KLGD equations:

$$\Gamma_P \frac{\partial P_3}{\partial t} + 2\alpha_p P + 4\beta_p P^3 + 2\chi P A^2 + 6\gamma_p P^5 - g\frac{\partial^2 P}{\partial z^2} = E_3, \qquad (1a)$$

$$\Gamma_A \frac{\partial A}{\partial t} + 2\alpha_a A + 4\beta_a A^3 + 2\chi P^2 A + 6\gamma_a A^5 - g\frac{\partial^2 A}{\partial z^2} = 0. \qquad (1b)$$

Eqs.(1) follow from the variation of KLGD free energy (see **Appendix A** for details). Here $\Gamma_{P,A}$ are the positive kinetic coefficients defining the Khalatnikov relaxation of the order parameters. The coefficients $\alpha_p = \alpha_T(T_P - T)$ and $\alpha_a = \alpha_T(T_A - T)$ change their sign at Curie temperature $T_P$ and AFE temperature $T_A$, respectively, $T$ is the absolute temperature, and $\alpha_T > 0$. The inequalities $\gamma_a > 0$, $\gamma_p > 0$ and $g > 0$ should be valid for the KLGD potential stability. The AFE-FE coupling coefficient $\chi$ should be positive for the stability of the AFE phase, and negative for the stability of the FE phase.

The boundary conditions for $P$ and $A$ at the film surfaces $z = 0$ and $z = h$ are of the third kind $\left(P \mp \Lambda_P \frac{\partial P}{\partial z}\right)\Big|_{z=0,h} = 0$ and $\left(A \mp \Lambda_A \frac{\partial A}{\partial z}\right)\Big|_{z=0,h} = 0$; they include extrapolation lengths $\Lambda_{A,P}$ [38, 39].

An equation for the surface charge is analogous to the Langmuir adsorption isotherm used in interfacial electrochemistry for adsorption onto a conducting electrode exposed to ions in a solution [40]. To describe the dynamics of surface ions charge density, we use a linear relaxation model,

$$\tau \frac{\partial \sigma}{\partial t} + \sigma = \sigma_0[\phi], \tag{2}$$

where the dependence of an equilibrium charge density $\sigma_0[\phi]$ on the electric potential $\phi$ is controlled by the concentration of surface ions, $\theta_i(\phi)$, at the interface $z = 0$ in a self-consistent manner, as proposed by Stephenson and Highland [25, 29]:

$$\sigma_0[\phi] = \sum_{i=1}^{2} \frac{eZ_i \theta_i(\phi)}{A_i} \equiv \sum_{i=1}^{2} \frac{eZ_i}{A_i} \left(1 + \rho^{1/n_i} \exp\left(\frac{\Delta G_i^{00} + eZ_i \phi}{k_B T}\right)\right)^{-1}, \tag{3}$$

where $e$ is an elementary charge, $Z_i$ is the ionization degree of the surface ions/electrons, $1/A_i$ are saturation densities of the surface ions. A subscript $i$ designates the summation on positive ($i = 1$) and negative ($i = 2$) charges, respectively; $\rho = \frac{p_{O2}}{p_{O2}^{00}}$ is the relative partial pressure of oxygen (or other ambient gas) [25], $n_i$ is the number of surface ions created per gas molecule. Two surface charge species exist, since the gas molecule had been electroneutral before its electrochemical decomposition started. The dimensionless ratio $\rho$ varies in a wide range from $10^{-6}$ to $10^6$ in the SH approach [25, 29].

Positive parameters $\Delta G_1^{00}$ and $\Delta G_2^{00}$ are the free energies of the surface defects formation at normal conditions, $p_{O2}^{00} = 1$ bar, and zero applied voltage $U = 0$. The energies $\Delta G_i^{00}$ are responsible for the formation of different surface charge states (ions, vacancies, or their complexes). Specifically, exact values of $\Delta G_i^{00}$ are poorly known even for many practically important cases, and so hereinafter they are regarded varying in the range ~(0 − 1) eV [25]. At that the difference $\Delta G_1^{00} - \Delta G_2^{00}$ can play a crucial role in the overall behavior of a ferroelectric film covered by the ions [25]. Notably, the developed solutions are insensitive to the specific details of the charge compensation process [41], but are sensitive to the thermodynamic parameters of corresponding reactions [42].

### III. FREE ENERGY OF ANTIFERRO-IONIC SYSTEM AND CALCULATION DETAILS
#### A. Free energy of the antiferroionic system

Since the stabilization of a single-domain polarization in ultrathin perovskite films takes place due to the chemical switching (see e.g. Refs.[13, 25, 29]), we a fortiori can assume the same for an AFE film. Thus, we will assume that the distributions of $P(x, y, z)$ and $A(x, y, z)$ do not deviate significantly from their values, averaged over the film thickness, which are further abbreviated as "**polarization**" $P \cong \langle P \rangle$ and "**anti-polarization**" $A \cong \langle A \rangle$. In this case, the behavior of the polarization $P$, anti-polarization $A$, and surface charge density $\sigma$ can be described via nonlinear coupled algebraic equations, similar to the ones derived in Refs. [30 - 34] for $P$ and $\sigma$.

Below we consider either the stationary case or adiabatic conditions, when $\sigma = \sigma_0$. The expression for the free energy density, $f = \frac{G[P,A,\Psi]}{S}$, which minimization gives the coupled equations for polarization dynamics, is the sum of the KLGD polar and antipolar ordering energy, $f_{AP}$, and the electrostatic energy $f_\Psi$. The energy $f_\Psi$ includes the polarization interaction energy with the overpotential $\Psi$, the energy of the electric field in the AFE film, and in the gap, correspondingly, and the surface charge energy. So that $f = f_{AP} + f_\Psi$, and the individual contributions are:

$$f_{AP} = h[\alpha_p P^2 + \alpha_a A^2 + \chi P^2 A^2 + \beta_p P^4 + \beta_a A^4 + \gamma_p P^6 + \gamma_a A^6], \tag{4a}$$

$$f_\Psi = -\Psi P - \varepsilon_0 \varepsilon_{33}^b \frac{\Psi^2}{2h} - \frac{\varepsilon_0 \varepsilon_d}{2} \frac{(\Psi - U)^2}{\lambda} + \int_0^\Psi \sigma_0[\varphi] d\varphi. \tag{4b}$$

Here $\alpha_p = \alpha_{pT}(T - T_P) + \frac{2g}{h\Lambda_P}$ and $\alpha_a = \alpha_{aT}(T - T_A) + \frac{2g}{h\Lambda_A}$ are the thickness-dependent and temperature-dependent functions. The terms $\sim g/h$ originate from "intrinsic" gradient-correlation size effects. The voltage $U$ is applied between the electrodes.

The free energy given by Eqs.(4) has an absolute minimum at high $\Psi$. According to the Biot's variational principle [43], we can further use a thermodynamic potential, which partial minimization over $P$ will give the coupled equations of state, and, at the same time, it has an absolute minimum at finite $P$ values.

### B. Calculation details

So, a formal minimization of Eqs.(4), $\frac{\partial f}{\partial P} = -\Gamma_P \frac{\partial P}{\partial t}, \frac{\partial f}{\partial A} = -\Gamma_A \frac{\partial A}{\partial t}$, and $\frac{\partial f}{\partial \Psi} = 0$, leads to the coupled time-dependent relaxation-type differential equations for the polarization and anti-polarization, and overpotential:

$$\Gamma_P \frac{\partial P}{\partial t} + 2(\alpha_p + \chi A^2)P + 4\beta_p P^3 + 6\gamma_p P^5 = \frac{\Psi}{h}, \tag{5a}$$

$$\Gamma_A \frac{\partial A}{\partial t} + 2(\alpha_a + \chi P^2)A + 4\beta_a A^3 + 6\gamma_a A^5 = 0, \tag{5b}$$

$$\frac{\Psi}{h} = \frac{\lambda(\sigma_0 - P) + \varepsilon_0 \varepsilon_d U}{\varepsilon_0 (\varepsilon_d h + \lambda \varepsilon_{33}^b)}. \tag{5c}$$

The overpotential $\Psi$ contains the contribution from surface charges proportional to $\sigma_0$, the depolarization field contribution proportional to $P$, and the external potential drop proportional to $U$.

The equations (5) were solved numerically, and obtained stationary solutions were substituted in the energy (4) in order to determine the energy of the corresponding state. Since Eq.(5b) is homogeneous, one can find the static solutions for $A$, namely $A = 0$ or $A^2 = \frac{2\beta_a \pm \sqrt{\beta_a^2 - 12\gamma_a(\alpha_a +)\chi P^2}}{6\gamma_a}$, and substitute them in Eqs.(5a). Further substitution of the overpotential $\Psi$ (as the function of $\sigma_0[U]$, $P$ and $U$) from Eq.(5c) to Eq.(5a) allows to plot the parametric dependence $U(P)$, which reversal gives us the static $P(U)$ and $A(U)$ curves.

The KLGD thermodynamic potential can be further expanded in $A$, $P$ and $\Psi$ powers, assuming that $|eZ_i\Psi/k_BT| \ll 1$ (see **Appendices B** and **C** [35] for details). In result we obtain the expression for the free energy:

$$F[P, A] = \alpha_{pR} P^2 + \beta_{pR} P^4 + \gamma_{pR} P^6 - (E_{SI} + E_{act})P + \chi_R P^2 A^2 + \alpha_a A^2 + \beta_a A^4 + \gamma_a A^6. \tag{6}$$

Renormalized coefficients, $\alpha_{pR}, \beta_{pR}, \gamma_{pR}$, and $\chi_R$, a built-in field $E_{SI}$ and an acting field $E_a$ are

$$\alpha_{pR}(T,\rho,h) = \alpha_p\big(1 + S(T,\rho,h)\big) + \frac{\lambda}{2\varepsilon_0(\varepsilon_d h + \lambda\varepsilon_{33}^b)}, \tag{7a}$$

$$\beta_{pR}(T,\rho,h) = \big(1 + S(T,\rho,h)\big)\beta_p, \qquad \gamma_{pR}(T,\rho,h) = \big(1 + S(T,\rho,h)\big)\gamma_p, \tag{7b}$$

$$\chi_R(T,\rho,h) = \big(1 + S(T,\rho,h)\big)\chi, \tag{7c}$$

$$E_{SI}(T,\rho,h) = \frac{\lambda}{\varepsilon_0(\varepsilon_d h + \lambda\varepsilon_{33}^b)}\sum_{i=1,2}\frac{eZ_i}{A_i}f_i(T,\rho), \qquad E_{act}(U,h) = -\frac{\varepsilon_d U}{\varepsilon_d h + \lambda\varepsilon_{33}^b}. \tag{7d}$$

The last term in Eq.(7a), $\frac{\lambda}{2\varepsilon_0(\varepsilon_d h + \lambda\varepsilon_{33}^b)}$, originated from the depolarization field. Since, as a rule, $\varepsilon_d h \gg \lambda\varepsilon_{33}^b$, the acting field is close to an external field, $E_{act} \approx -\frac{U}{h}$. Also, we introduce positive functions in Eq.(7):

$$S(T,\rho,h) = \frac{\lambda h}{\varepsilon_0(\varepsilon_d h + \lambda\varepsilon_{33}^b)}\sum_{i=1,2}\frac{(eZ_i f_i(T,\rho))^2}{A_i k_B T}, \quad f_i(T,\rho) = \left(1 + \exp\left(\frac{\Delta G_i^{00}}{k_B T}\right) + \rho^{1/n_i}\right)^{-1}. \tag{7e}$$

### B.1. The case of the second order phase transition (2-4 Landau expansion)

Note that in the case of the AFE film with the second order phase transitions, when $\beta_{a,p} > 0$ and $\gamma_{a,p} = 0$, analytical solutions for the phase energy, order parameters and critical (or coercive) fields of double (or single) hysteresis loops are possible after a trivial minimization of the Landau energy (6). They are summarized in **Table I.** The boundaries between the AFE phase, mixed ferrielectric FEI phase, pressure-induced FE-like AFI phase, and paraelectric phase correspond to the condition of the phase energies equality. Below we will be specifically interested in the pressure-induced transition from the AFE phase to the FE-like AFI phase. The FE-like AFI phase becomes absolutely stable if $\alpha_{pR} < 0$, and $F_{AFI} < 0$ is minimal in comparison with $F_{AFE}$ and $F_{FEI}$. When $\alpha_a > 0$ at $T > T_A$ and we put $\chi < 0$ for the AFE phase stability, the condition

$$\alpha_{pR}(T_A, \rho, h) \leq 0 \tag{8}$$

is sufficient for the absolute stability of the AFI phase.

**Table I.** Thermodynamic phases, order parameters, and critical fields for $\beta_{a,p} > 0$ and $\gamma_{a,p} = 0$.

| Phase | Order parameters at $E_{act} = E_{SI}$ | Free energy at $E_a = E_{SI}$ | Critical field(s) $E_c$ |
|---|---|---|---|
| AFE | $A_S = \pm\sqrt{-\alpha_a/2\beta_a}$<br>$P_S = 0$<br>$\Delta P \cong \pm\sqrt{-\frac{\alpha_{pR}}{6\beta_{pR}}}$<br>(see **Fig. 1b** for definitions) | $F_{AFE} = -\frac{\alpha_a^2}{4\beta_a},$<br>$\alpha_a < 0$ | $E_{c1,2} = \frac{\pm 1}{3\sqrt{3}}\frac{\left(-\alpha_{pR}+\frac{\chi_R\alpha_a}{2\beta_a}\right)^{3/2}}{\beta_{pR}-\frac{\chi_R^2}{4\beta_a}} - E_{SI}$<br>$E_{c3,4} = \pm\frac{1}{3\sqrt{3}}\frac{(-\alpha_{pR})^{3/2}}{\beta_{pR}} - E_{SI}$ |
| FEI | $P_S = \pm\sqrt{\frac{2\alpha_{pR}\beta_a - \chi_R\alpha_a}{4\beta_a\beta_{pR} - \chi_R^2}}$<br>$A_S = \pm\sqrt{\frac{2\alpha_a\beta_{pR} - \chi_R\alpha_{pR}}{4\beta_a\beta_{pR} - \chi_R^2}}$ | $F_{FEI} =$<br>$-\frac{\alpha_a^2\beta_{pR} + \alpha_{pR}^2\beta_a - \chi_R\alpha_a\alpha_{pR}}{4\beta_a\beta_{pR} - \chi_R^2}$ | $E_{c1,2} = \frac{\pm 1}{3\sqrt{3}}\frac{\left(-\alpha_{pR}+\frac{\chi_R\alpha_a}{2\beta_a}\right)^{3/2}}{\beta_{pR}-\frac{\chi_R^2}{4\beta_a}} - E_{SI}$ |
| FE-like AFI | $P_S = \pm\sqrt{-\frac{\alpha_{pR}}{2\beta_{pR}}}$, $A_S = 0$ | $F_{AFI} = -\frac{\alpha_{pR}^2}{4\beta_{aR}}$, $\alpha_{pR} < 0$ | $E_{c1,2} = \pm\frac{1}{3\sqrt{3}}\frac{(-\alpha_{pR})^{3/2}}{\beta_{pR}} - E_{SI}$ |
| PE | $A_S = P_S = 0$ | 0 | absent |

We use a Gaussian Process model (**GPM**) for rapid exploration and prediction of phase diagrams and order parameters corresponding to the free energy (6) for the case $\beta_{a,p} > 0$ and $\gamma_{a,p} = 0$ in **Appendix D** [35]. The material parameters are listed in **Table D1**. Results are shown in **Figs. D1-D4** [35]. In the GP predicted image, we can see similar distinctive region with different color coding and thus gave us the interpretation of individual phases. We can have better prediction of phases with more advanced acquisition function (exploration-exploitation) to sample as per user criteria (for example- we can conduct adaptive sampling where the objective function is higher). We observed that as $\rho$ increase or decreases by the order of 10 for temperature $T < T_p$, we are approaching towards FEI phase where with further increase or decreases of $\rho$ by the order of 10, the deeper wells for order parameter $A$ are shifted to deeper wells for order parameter $P$. Also, as the parameters $h$ and $\Delta G_i^{00}$ decreases, the FEI region shrinks and expands respectively for the same parameter space of $T, \rho$. As per defined in Table I, we did not find the AFI region ($A = 0, P_S \neq 0$) for any of **Figures D1-D4**. The codes are available at: https://colab.research.google.com/drive/1jKbf2Yo5Y_ezmHaadjTB6qtxXYVfzg9T

### B.2. The case of the first order phase transition (2-4-6 Landau expansion)

For the illustration of numerical results, we use an AFE film with the thickness $h = (5 - 50)$ nm separated from the tip electrode by the gap of a thickness $\lambda = (0 - 2)$ nm. Also, we regard that the ion formation energies are equal at normal conditions and small, $\Delta G_1^{00} = \Delta G_2^{00} = 0.2$eV. Using results of Haun et al. [44], we determine KLGD expansion coefficients for a model antiferroelectric PbZrO$_3$ (see **Appendix C** [35]). It appeared that $\beta_{a,p} < 0$ and $\gamma_{a,p} > 0$ for this material, and so 2-4-6 KLGD expansion must be used. Corresponding expansion coefficients and other parameters are listed in **Table C1** in **Appendix C** [35].

However, the coefficients from **Table C1** do not capture all important features of a bulk PbZrO$_3$, as we note several discrepancies in the temperature behavior of the calculated here and measured critical fields and polarization hysteresis shape [45, 46, 47, 48]. We relate the discrepancy with the strong influence of antiferrodistorsive subsystem (oxygen octahedra tilt) that is missed in the used KLGD model. The complete and self-consistent phenomenological description of PbZrO$_3$ antiferroelectric and electrocaloric properties [49], and anomalous ferroelectricity [50] allowing for an antiferrodistortion is an important problem beyond the scope of this work. However, the ferroelectricity observed in thin PbZrO$_3$ films [51, 52] is in scope of our work, at least partially.

When calculating the hysteresis loops of polarization $P(U)$ at nonzero frequency $\omega$ of the applied voltage, $U = U_0\sin(\omega t)$, we use different kinetic coefficients $\Gamma_P \gg \Gamma_A$ to make the characteristic relaxation time of $P$ much longer than the relaxation time of $A$. The strong inequality $\Gamma_P \gg \Gamma_A$ leads to the polarization relaxation with the characteristic time $\tau = \frac{\Gamma_P}{2|\alpha_p|}$, while the anti-polarization behaves adiabatically. Hence the dimensionless frequency $w = \omega\tau$ govern the polarization response to external field.

To avoid "sticking" of the system at local minima, we applied a very small fictious "antifield" (with an amplitude smaller than $10^{-6}$ V/nm) acting on the anti-polarization $A$. The physical origin of the antifield is related with variations of the local electric field. The calculations were performed and visualized in Mathematica 12.2 software (https://www.wolfram.com/mathematica) and available at https://notebookarchive.org/2021-06-bislk1z.

## IV. RESULTS

### A. A phase diagram, free energy relief and hysteresis loops of a thick antiferroelectric film without surface ions

The phase diagram of a thick AFE film as function of temperature is shown in **Fig. 2a.** Here the pressure-induced surface ions are absent and the gap is very small to put the film close to bulk conditions. In accordance with available experimental results the diagram contains the region of a pure AFE phase followed by an AFE phase coexisting with a weak FE phase, and then by a paraelectric (**PE**) phase. The first order phase transition between the AFE and the PE phase occurs at $T_A \cong 490$ K, and the boundary between the AFE and the AFE-FE phases is very diffuse and located around $T_P \cong 460$ K.

Contour maps of the free energy dependence on the polarization and anti-polarization are shown in the upper insets **(b)-(d)** to **Fig. 2** for different temperatures $T =$ 200, 480 and 500 K, and $E = 0$. There are two relatively deep $A$-wells ($A = \pm A_S, P = 0$) and two very shallow $P$-wells ($P = \pm P_S, A = 0$) in the AFE-FE coexisting region [inset **(b)** for 200 K]. Both of $A$-wells, as well as both of $P$-wells, have the same depth. The $A$-wells and $P$-wells are separated by the four saddles. The potential wells (especially $A$-wells) become shallower with the temperature increase up to $T_A$ [inset **(c)** for 480 K]. At temperatures lower that $T_A$ the potential wells have negative energy, i.e., they correspond to the coexisting stable polar and anti-polar states. Their energy become positive and corresponding states become metastable at $T > T_A$ [inset **(d)** for 500 K]. The wells disappear in a deep PE phase.

The relief of the free energy determines the temperature behavior of polarization dependence on the external $E$-field, $P(E)$, where $E = \frac{U}{h}$. The quasi-static dependence $P(E)$ is hysteretic, and its shape gradually changes from a single ferroelectric-type loop to a loop with constriction, and then to a double antiferroelectric-type loop with the temperature increase from 200 K to 500 K, respectively [see red loops in the insets **(e)-(g)** to **Fig. 2**]. However, the transition from the single to a double loop takes place only when $E > E_{cr}$, where $E_{cr}$ is the static critical field (see dark-red parts of the loops). The dependence $P(E)$ is quasi-linear at $E < E_{cr}$ [see red parts of the $P(E)$ curves]. Notable that $P(E)$ are antisymmetric with respect to the $E$-axis, since the built-in field is absent due to the absence of surface ions.

Note, that the phase set (AFE, AFE-FE and PE) calculated within 2-4-6 Landau expansion with material parameters from **Table C1** [35] and shown in **Fig. 2a** differs from the analogous set (AFE, FEI, PE) calculated within 2-4 Landau expansion with material parameters from **Table D1** and shown in **Fig. D1-D4** [35]. Since $\beta_{a,p} > 0$ and $\gamma_{a,p} = 0$ and $\beta_{a,p} < 0$ and $\gamma_{a,p} > 0$ for the 2-4 and 2-4-6 Landau expansions, respectively, as well as the effect of $\rho$ is not considered in **Fig. 2a**, the difference in the phase set seems natural.

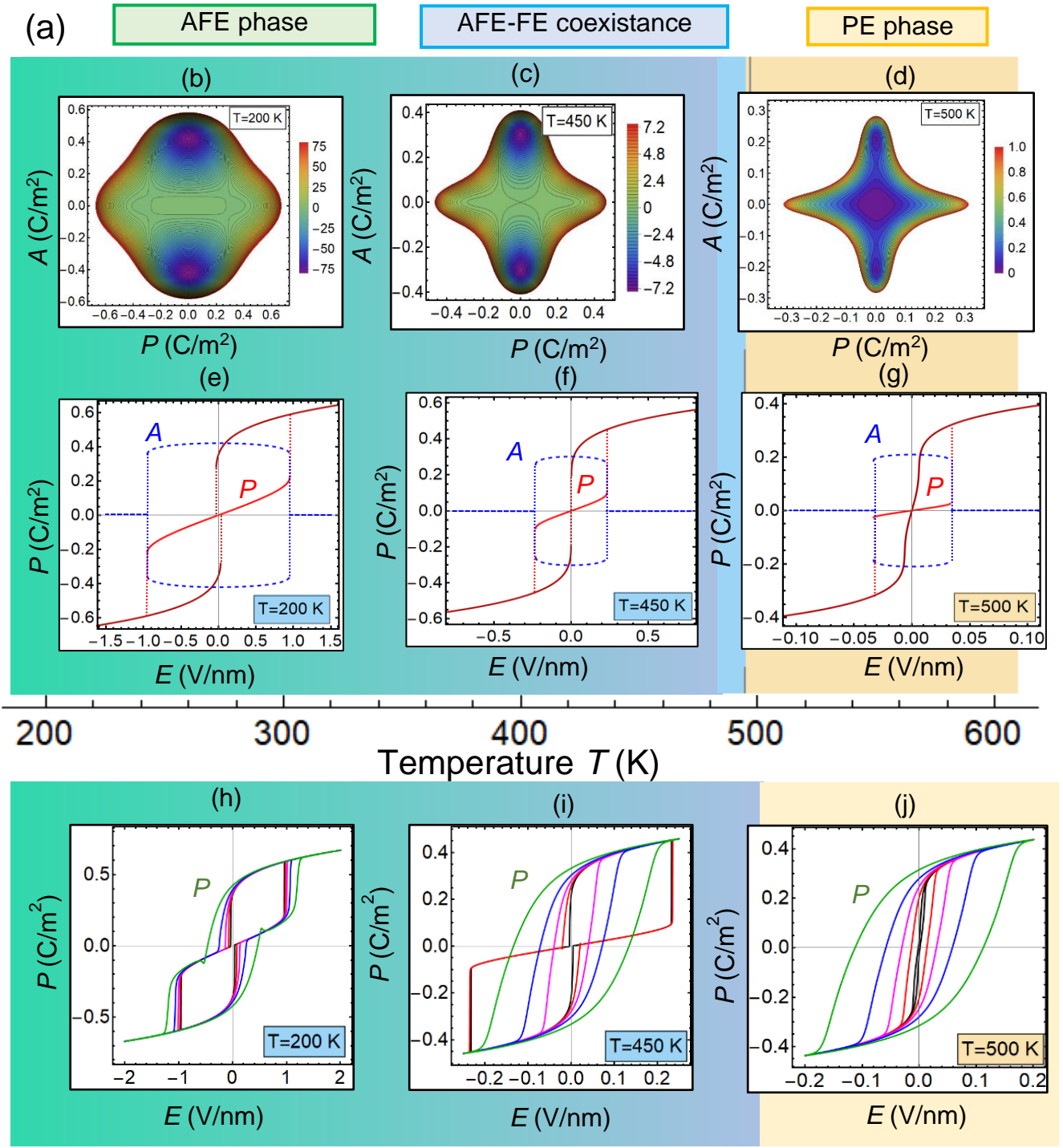

**FIGURE 2.** Phase diagram of an AFE film **(a)**. Contour maps of the free energy dependence on the polarization ($P$) and anti-polarization ($A$) for the temperature $T = 200$ K **(b)**, 450 **(c)** and 500 K **(d)**, and $E = 0$. Relative units are used for the energy color scale. Dependences of the anti-polarization $A$ (dashed and dotted blue loops) and polarization $P$ (dashed red and dark-red loops) on the static external electric field $E$ calculated for temperature $T = 200$ K **(e)**, 450 K **(f)**, and 500 K **(g)**, respectively. Polarization hysteresis $P(E)$ calculated for dimensionless frequencies $w = 0.3$ (black loops), 3 (red loops), 10 (magenta loops), 30 (blue loops), and 100 (green loops), and temperatures $T = 200$ K **(h)**, 450 K **(i)**, and 500 K **(j)**, respectively. For the plots **(h)-(j)** external electric field is $E = \frac{U_0}{h}\sin(\omega t)$, and dimensionless frequency $w = \frac{\omega \Gamma_P}{2|\alpha_p|}$. The pressure-induced surface charges are absent. Gap thickness $\lambda \leq 0.2$ nm, film thickness $h = 50$ nm. Other parameters are listed in **Table C1** [35].

Polarization hysteresis loops, $P(E)$, calculated for low dimensionless frequencies $w \ll 1$, almost coincide with the static $P$-curves [compare black loops in the plots **(h)-(j)** with red curves in the plots **(e)-(g)**]. The frequency increase in the range $1 \leq w \leq 3$ leads to the appearance of a thin constriction between the double loops in the AFE phase, as well as to the loop opening in the PE phase [see red loops in the plots **(h)-(j)**]. The width of the constriction increases significantly with further increase of $w$, and the loop acquires a ferrielectric-like shape in the AFE phase below 350 K [see magenta, blue and green loops in the plot **(h)**]. At the same time the frequency increase ($3 \leq w \leq 10$) at higher temperatures ($350$ K $< T < 450$ K) leads to an opening of a ferroelectric-like loop, meaning that the opening is a purely dynamic effect [see magenta, blue and green loops in the plot **(i)**]. The dynamic effect allows us to classify the temperature region as AFE-FE coexisting region. The high frequency $w \geq 10$ opens a pseudo-ferroelectric loop even in the PE phase [see magenta, blue and green loops in the plot **(j)**].

### B. The influence of surface ions on the phase diagram, free energy relief and hysteresis loops of antiferroelectric films

The free energy as a function of polarization ($P$) and anti-polarization ($A$) calculated for several temperatures $T$ and relative partial oxygen pressures $\rho$, which values are listed for each column/row, is shown in **Fig. 3**. There are two very deep $A$-wells ($A = \pm A_S$, $P = 0$) and two very shallow $P$-wells ($P = \pm P_S$, $A = 0$) in the deep AFE phase at the temperatures well below $T_P$ and $10^{-6} < \rho < 10^4$ [see columns **(a)-(c)** for $(200 - 400)$ K]; and the wells become shallower at $T \to T_A$ and eventually disappear with the temperature increase well above $T_A$ [see the column **(d)** for 500 K]. Both of $A$-wells have the same depth independently on the values of $\rho$ and $T$. The relative depth of $P$-wells depends on $\rho$ and $T$ values. The right well ($P = +P_S$) is evidently deeper for $\rho \gg 1$ and $T > T_A$; they are almost equal for $10^{-4} < \rho < 10^4$ and $T < 400$ K; and exactly equal for $\rho = 1$ independently on $T$; and the left well ($P = -P_S$) becomes evidently deeper for $\rho \ll 1$ and $T > T_A$ [compare the top, middle and bottom rows]. The origin of the $P$-wells asymmetry is the built-in electric field, $E_{SI}$, induced by the surface ions charge. This field is approximately proportional to the sum ($\rho^{n_i} - \rho^{-n_i}$), and thus the magnitude of $E_{SI}$ increases with the deviation of $\rho$ from unity, and the sign of $E_{SI}$ changes when $\rho \to \frac{1}{\rho}$, that is a direct consequence of Eqs.(7d) and (7e) at $\Delta G_1^{00} = \Delta G_2^{00}$. The sign change of $E_{SI}$ explains the left (or right) asymmetry of the $P$-wells.

The $A$-wells and $P$-wells are separated by the four saddles. At temperatures lower that $T_A$ both $A$-wells and $P$-wells have negative energy, i.e., they correspond to stable polar and anti-polar states. The energy of $A$-wells become positive and corresponding states become metastable at $T > T_A$ [see the column **(d)** for 500 K]. The $A$-wells disappear at $T \gg T_A$. Both $P$-wells acquire positive energy at $T > T_P$ and disappear at $T \gg T_A$ only when $10^{-2} < \rho < 10^2$. Either right or left negative well appears when $\rho > 10^2$ or $\rho < 10^{-2}$, respectively. This means that the built-in field induced by the surface ions charge creates and supports an FE-like AFI state, that is characterized by the asymmetric potential relief, and the asymmetry increases with excess ($\rho \gg 1$) or deficiency ($\rho \ll 1$) of oxygen ions at the film surface.

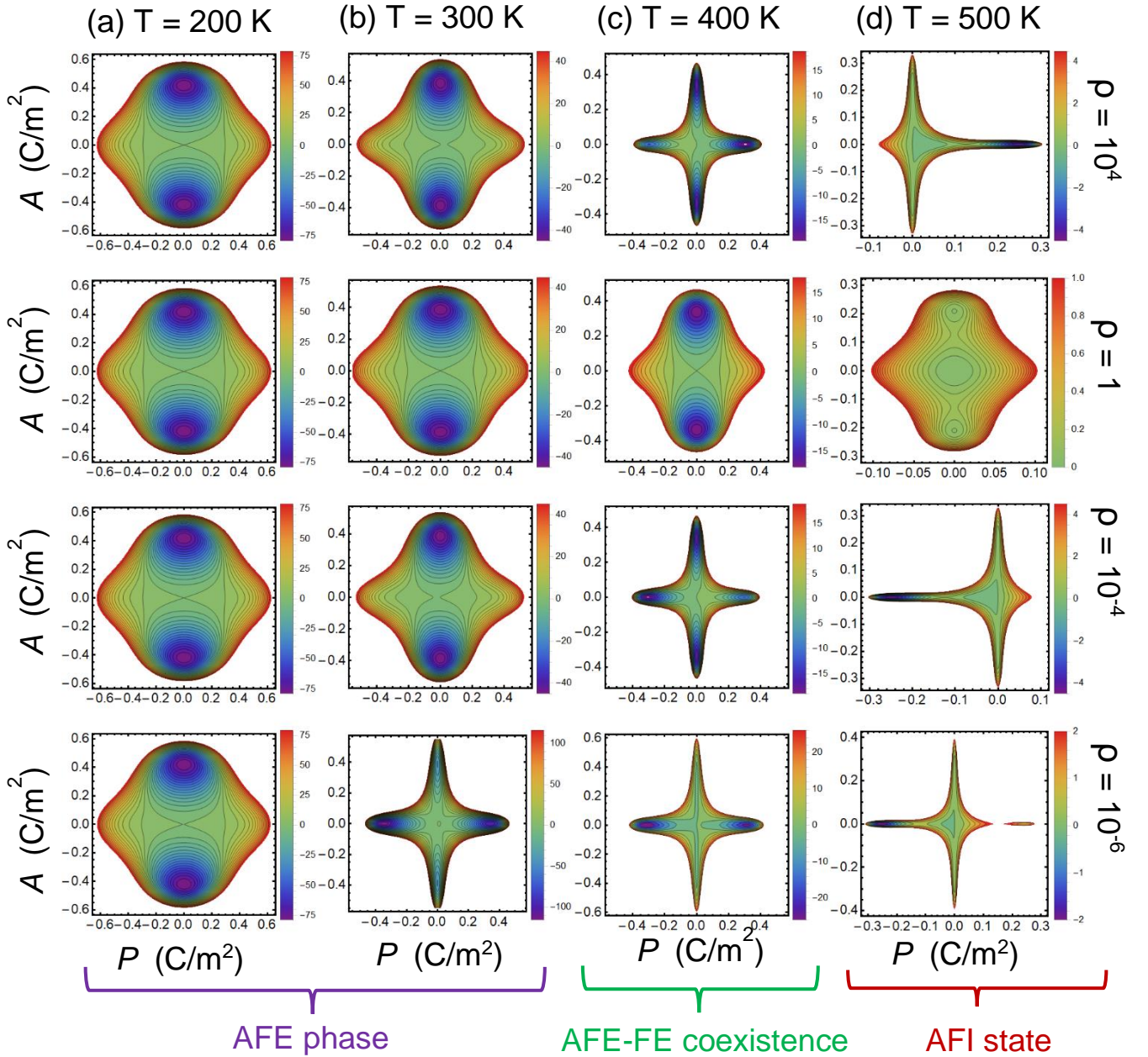

**FIGURE 3.** The free energy as a function of polarization ($P$) and anti-polarization ($A$) calculated for the temperature $T = 200$, 300, 400 and 500 K [columns **(a)**, **(b)**, **(c)** and **(d)**] and relative partial oxygen pressures $\rho = 10^4, 1, 10^{-4}, 10^6$, which values are listed for each column/row. Relative units are used for the energy color scale. An external electric field is absent, $E = 0$. The thickness of AFE film is $h = 50$ nm, gap thickness $\lambda = 2$ nm, and ion formation energies $\Delta G_1^{00} = \Delta G_2^{00} = 0.2$ eV.

Since the case $\rho \gg 1$, corresponding to high oxygen excess, is hard to realize in practice, below we mainly discuss the easy realizable case of the oxygen deficiency, $\rho \leq 1$, keeping in mind that the physical picture at $\rho \geq 1$ differs the one at $\rho \leq 1$ only by the horizontal asymmetry (the left or the right well is deeper) of the potential relief, and corresponding built-in fields leading to the left or right shift of polarization loops, respectively.

The dependences of the anti-polarization $A$ (blue curves) and polarization $P$ (red and dark-red curves) on the static electric field $E$ are shown in **Fig. 4** for temperatures $T = (200 – 500)$ K (columns **a-d**) and relative oxygen pressures $\rho = 1, 10^{-4}, 10^{-6}$ (from the top to the bottom rows), which values are listed for each column/row. The red and dark-red polarization curves are two nonzero stable solutions of Eq.(5a). Dotted vertical lines show the thermodynamic transitions between different polar and anti-polar states.

Changes that occur with the shape of the loop in vertical columns **(a, b, c)** and **(d)**, with an increase in temperature, mainly consist in a decrease in the width and height of $A$ and $P$ loops, and the appearance of a noticeable horizontal asymmetry of $P$-loops with a change in the relative oxygen pressure $\rho$ from 1 to $10^{-6}$.

The features of $A(E, T)$ behavior depend on T and $\rho$ much weaker than the features of $P(E, T)$. The $A$-loop is absent only for $\rho < 10^{-4}$ and above $(350 − 450)$ K, which indicate the transition to the FE-like AFI state due to the absence of "ionic support" of the AFE state. At lower partial oxygen pressures, the shape of $A$-loop is close to rectangular, its width and height depend on temperature and pressure. This rectangle abruptly "collapses" to two zero lines at external fields larger than the critical value $E_c$, since the anti-polarization $A$ no longer exists. This corresponds to a transition from the state with an antiparallel configuration of spontaneous dipoles to the state with their parallel orientation in both sublattices.

A decrease of $A$ magnitude with increasing an electric field $E$ in the range $E < E_c$ corresponds to an increase in $P$, and the $A(E)$ curves are almost symmetric with respect to the $E$-axis, while the $P(E)$ curves acquire a more noticeable horizontal asymmetry with the increase of $\rho$ from 1 to $10^{-6}$.

$P$-curves supplemented with red dotted vertical lines are virtually static double loops at $\rho = 1$ and become hysteresis-less only at $T > 500$ K. The coercive field and the loop height decrease with temperature at $\rho = 1$ (as it should be). The double $P$-loops transforms to the loops with a thin constriction with $\rho$ decrease and temperature increase. Then the constriction significantly increases for $\rho = 10^{-4}$ and $T \geq 300$ K, and eventually the static loop with a wide constriction transforms to a ferroelectric-like single loop for $\rho = 10^{-6}$ and $T \geq 400$ K. The FE-like single $P$-loops correspond to the AFI state, because the surface ions support the FE polarization, and suppress the antipolar order.

Note that the $P$-curves have no horizontal asymmetry for $\rho = 1$, since the built-in field $E_{SI}$ is absent in the case. The $P$-curves are slightly right-shifted for $\rho < 1$, since $E_{SI}$ is positive. The shift increases significantly with $\rho$ decrease below $10^{-4}$ and $T$ increase above 400 K. The $P$-curves for $\rho = 10^{4}$ are left-shifted, since $E_{SI}$ is negative in the case; they are shown in **Fig. E1** in **Appendix E** [35].

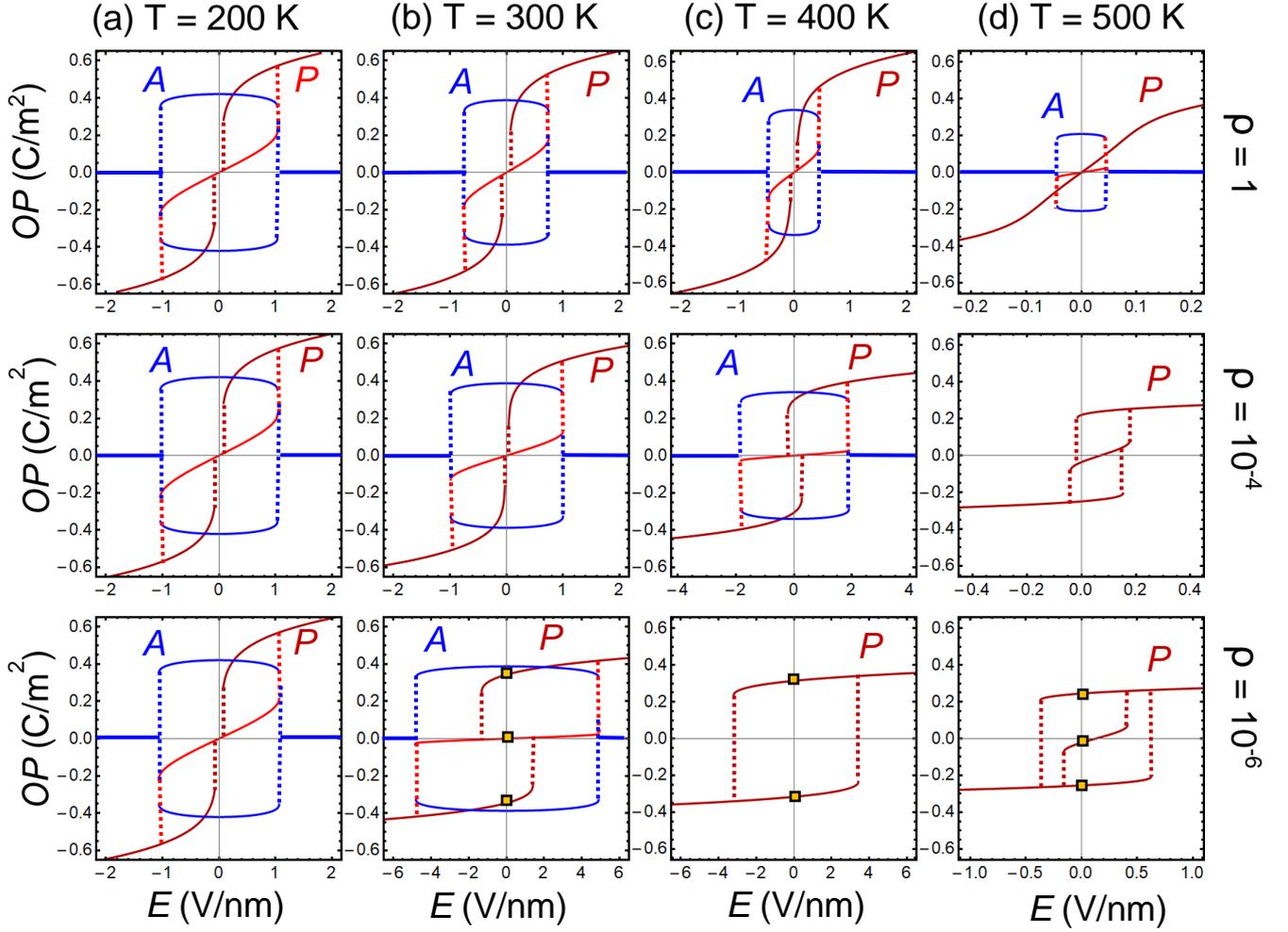

**FIGURE 4.** Static dependences of the order parameters (OP) – anti-polarization $A$ (solid blue curves) and polarization $P$ (solid red and dark-red curves) on external electric field $E$ calculated for temperature $T = 200, 300, 400$ and $500$ K [columns **(a)**, **(b)**, **(c)** and **(d)**] and relative partial oxygen pressures $\rho = 1, 10^{-4}, 10^{-6}$, which values are listed for each column/row. Dotted vertical lines show the thermodynamic transitions between different polar and antipolar states; yellow rectangles mark 2 or 3 stable polarization states. An external electric field is $E = \frac{U}{h}$. Other parameters are the same as in **Fig. 3.**

Polarization hysteresis loops $P(E)$ calculated for dimensionless frequencies $w = 0.3 - 100$, temperatures $T = (200 - 500)K$ and relative oxygen pressures $\rho = 1 - 10^{-6}$ are shown in **Fig. 5**. Under the pressure decrease from 1 to $10^{-4}$ and relatively low frequencies $w \leq 10$ the loop shape demonstrates a continuous transition from a double loop in the AFE phase to a single loop in the AFI state, and then the loop disappears in a deep PE phase. Only a single loop exists at $\rho = 10^{-6}$ and $T \geq 300$ K, and it gradually degrades to a shifted hysteresis-less PE curve with a temperature increase far above 600 K (that is not shown in the figure). The frequency increase ($w \geq 30$) transforms a double loop to a loop with constriction, and then it "opens" a single loop. The loop opening is a dynamic effect similar to the one, shown in **Figs. 2h-j**. Both quasi-static ($w \leq 10$) and dynamic ($w \geq 30$) loops are slightly right-shifted at $\rho < 1$, and the shift is proportional to the built-in field $E_{SI}$. The shift increases with $\rho$ decrease below $10^{-4}$ and $T$ increase above 400 K. The $P$-curves for $\rho = 10^4$, which are left-shifted (since $E_{SI}$ is negative in the case), are shown in **Fig. E2** in **Appendix E**.

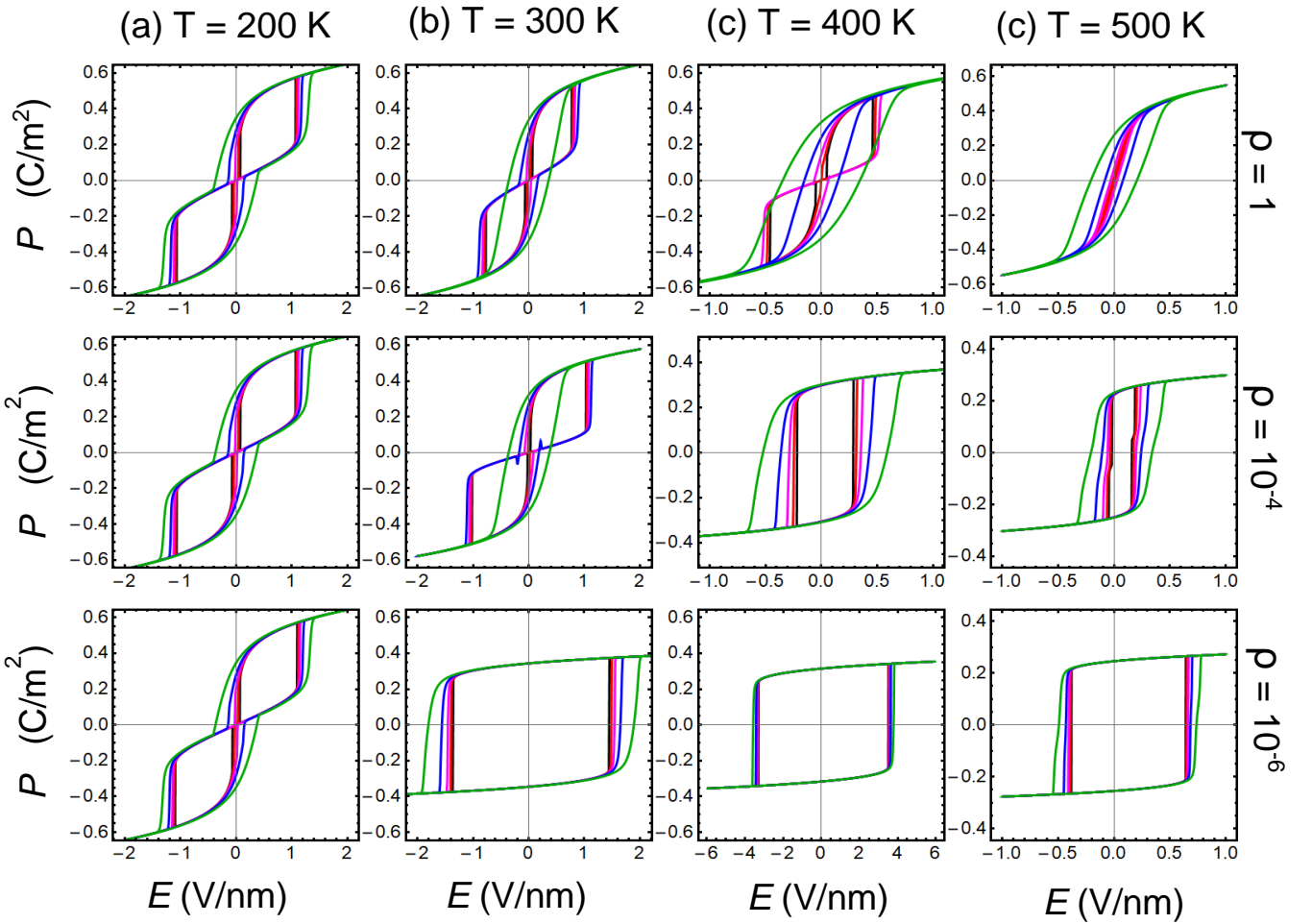

**FIGURE 5.** Polarization hysteresis loops, $P(E)$, calculated for dimensionless frequencies $w = 0.3$ (black loops), 3 (red loops), 10 (magenta loops), 30 (blue loops), and 100 (green loops), temperatures $T = 200$ K, 300 K, 400 K and 500 K [columns **(a)**, **(b)**, **(c)** and **(d)**, respectively], and relative partial oxygen pressures $\rho = 1, 10^{-4}, 10^{-6}$, which values are listed for each column/row. External electric field is $E = \frac{U_0}{h}\sin(\omega t)$, and $w = \frac{\omega \Gamma_P}{2|\alpha_p|}$. Other parameters are the same as in **Fig. 3**.

A typical phase diagram of a thin AFE film in dependence on the temperature $T$ and relative oxygen pressure $\rho \leq 1$ in shown in **Fig. 6a** for $h = 50$ nm and in in **Fig. 6b** for $h = 5$ nm. There are an AFE phase, an AFE coexisting with a weak FE phase, a FE-like AFI phase, and an electret-like PE phase. The insets **(c, d, e)** illustrate the typical free energy relief is the AFE-FE, FE-like FEI and electret-like PE phase, respectively. The insets **(f, g, h)** are typical hysteresis loops in these phases. The phase diagram plotted for the relative pressures from $10^{-6}$ to $10^6$ is shown in **Fig. E3** in **Appendix E** [35].

The wide light-blue AFE FE coexistence region increases with $h$ decrease (compare **Fig. 6a** and **Fig. 6b**), but it is located at the temperatures lower than 500 K (that is slightly higher than $T_A \cong 490$ K) independently on the film thickness, because the anti-polarization is insensitive to the depolarization field. The relatively small light-green region of the AFI phase decreases significantly with $h$ decrease (compare **Fig. 6a** and **Fig. 6b**), because the polarization is very sensitive to the depolarization field $E_d$, which is inversely proportional to the film thickness, $E_d \sim -\frac{\lambda P}{\varepsilon_0(\varepsilon_d h + \lambda \varepsilon_{33}^b)}$ [see Eq.(5c)]. The AFI phase corresponds to rather low relative oxygen pressures ($\rho \leq 2 \cdot 10^{-4}$ for $h = 50$ nm or $\rho \leq 10^{-5}$ for $h = 5$ nm), but exist in a relatively wide temperature range (250 K $\leq T \leq$ 500 K for $h = 50$ nm, or 310 K $\leq T \leq$ 475 K for $h = 5$ nm). The boundary between the AFE-FE region and the AFI phase is close to a rounded corner. The boundary between the AFE-FE, the AFI and the PE phase (local inside a sand-colored region) is close to the vertical line $T \cong 500$ K.

Note, that the phase set, namely AFE, AFE-FE, FE-like AFE, and electret-like PE, shown in **Figs. 6** differs from the analogous set (AFE, FEI, PE) shown in **Fig. D1-D4** [35]. The difference originated from the difference in Landau expansion coefficients, $\beta_{a,p} > 0$ and $\gamma_{a,p} = 0$ for 2-4 Landau expansion, apart $\beta_{a,p} < 0$ and $\gamma_{a,p} > 0$ for the 2-4-6 Landau expansion (compare material parameters from **Table C1** with material parameters from **Table D1**[35]).

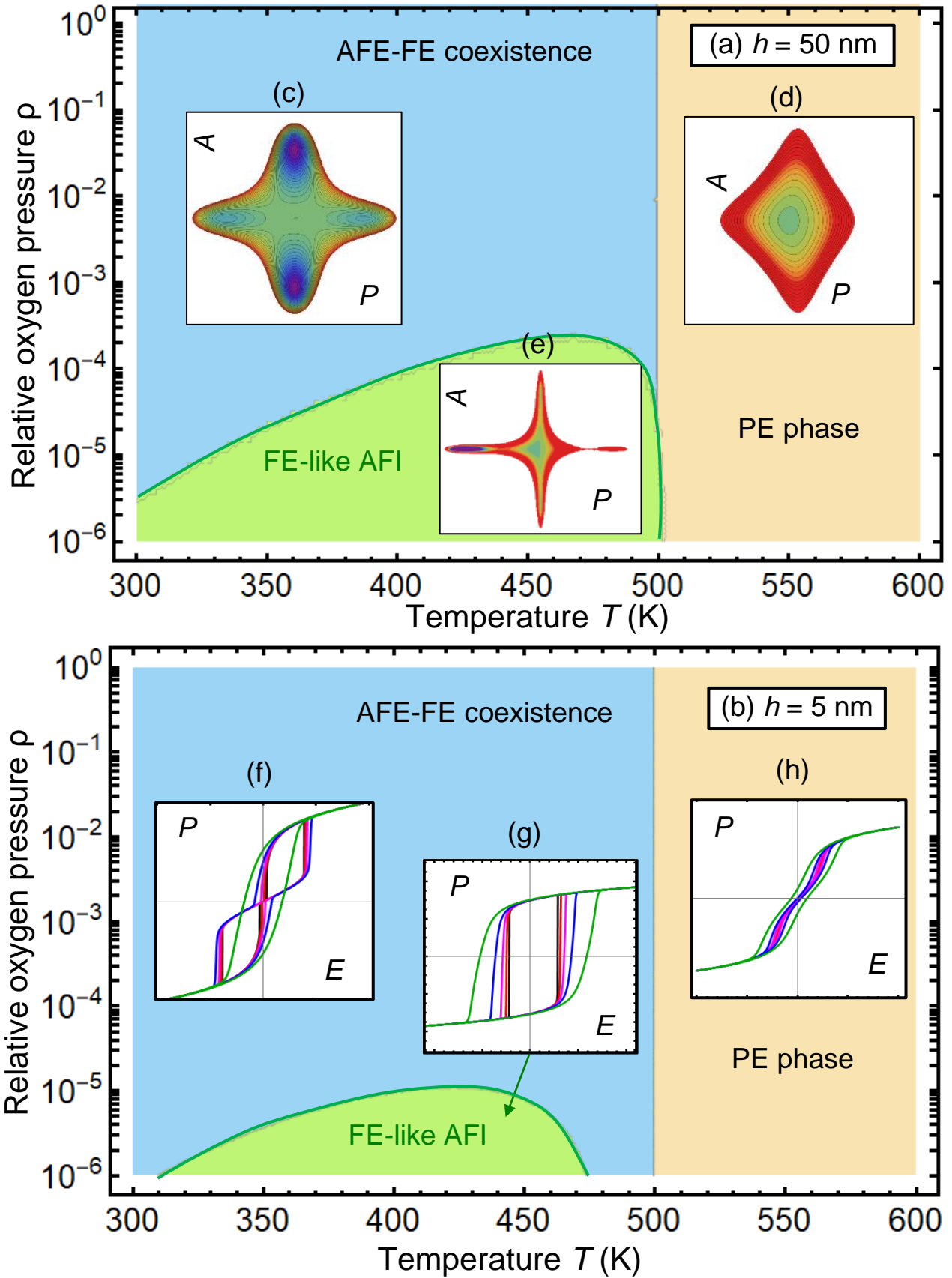

**FIGURE 6.** Typical phase diagrams of thin AFE films with thickness $h = 50$ nm **(a)** and $h = 5$ nm **(b)** in dependence on the temperature $T$ and relative partial oxygen pressure $\rho$. There are an AFE phase coexisting with a weak FE phase, a FE-like AFI phase, and an electret-like PE phase. Typical free energy maps at $E = 0$ **(b-d)** and polarization

hysteresis loops $P(E)$ **(e-g)**. The description of the insets is the same as in **Figs. 3** and **5**, respectively. Other parameters are the same as in **Fig. 3.**

### C. The influence of surface ions on the energy and information storage in thin AFE films

It is well-known that an energy loss is an area inside a hysteresis loop $P(E)$, further abbreviated as **LA**, and a stored energy is equal to an area above the loop, further abbreviated as **SA**. The nonvolatile information storage requires high remanent polarization and not very small coercive bias, and, thus the optimal loop area **LA**. If the loop is absent and $P \cong \varepsilon E$, the loss is absent, and the stored energy is $\frac{\varepsilon}{2}E^2$. However, the nonvolatile information storage is impossible in the case.

For a single quasi-rectangular ferroelectric hysteresis loop with polarization saturation above the coercive field the stored energy is small (zero in the limit of a rectangular loop), and the loop area is given by the approximate expression $LA = P_S(E_{c2} - E_{c1})$ (see **Fig. 1b**). The expressions for the spontaneous polarization $P_S$ and coercive fields $E_{ci}$ in the case of the second order phase transitions are given in **Table I**. The area of a double loop is given by approximate expression $LA = \Delta P(E_{c3} - E_{c1}) + \Delta P(E_{c2} - E_{c4})$, and the stored energy is $SA = \frac{\varepsilon}{2}(E_{c1}^2 + E_{c2}^2) + \Delta P(E_{c1} - E_{c2})$ for the case of polarization saturation above the coercive fields a rough approximation $\frac{LA}{SA} \sim \frac{E_{c3} - E_{c4}}{E_{c1} - E_{c2}} - 1$ is valid. If all four coercive field exists, we obtain that $\frac{LA}{SA} \sim \left(1 - \frac{\chi_R \alpha_a}{2\alpha_{pR}\beta_a}\right)^{-3/2} - 1$ for the AFE film with the second order phase transitions [$\beta_{a,p} > 0$ and $\gamma_{a,p} = 0$ in Eq.(6)].

It follows from the above expressions that one need a large step-like (or quasi-linear with a small slope) hysteresis-less region (e.g., $P \cong \varepsilon E$, where $\varepsilon$ is small) for the maximal energy storage. Without the region the loop shape optimization for the energy storage in the space of parameters $\{\rho, T, h\}$ leads to a trivial result - no loop at all, i.e., to the hysteresis-less curve in the PE phase. The presence of the linear region adds additional trapezoidal area above the $P(E)$ curve, and the energy storage becomes favorable in the AFE phase also (see **Fig. 1b**). Since we have shown that the oxygen deficiency (or excess) can transform a double AFE loop to a single FE-like one (compare loops in **Fig. 5**), the usage of AFI film for the energy storage can be not beneficial, at the same time, the information storage in AFI films can possess several advantages. Let us discuss the question in more details for the AFE film with the first order phase transitions [$\beta_{a,p} < 0$ and $\gamma_{a,p} > 0$ in Eq.(6)].

**Figure 7a** illustrates that the double hysteresis loops with a pronounced linear region between the loops can exist at room temperature only in ultra-thin AFE films ($h \leq 10$ nm) if the relative oxygen pressure does not deviate significantly from the normal conditions, $10^{-3} \leq \rho \leq 1$. At the same time the linear region is almost absent for thicker films (see **Fig. 7b**). At the temperatures above $T_A$ an ultra-thin film becomes either paraelectric for $\rho \cong 1$, or electret-like for $\rho \ll 1$, and corresponding hysteresis-less $P(E)$ curves cannot be used for the information storage, because their area is zero (see **Fig. 7c**), while double hysteresis loops with a very small loop area (and, consequently, very small losses) can exist in a thicker film for $\rho \leq 10^{-4}$ and $T \geq T_A$ (see **Fig. 7d**). Similar plots for the film thickness 5, 10, 20 and 50 nm are shown in **Fig. E4** in **Appendix E** [35].

Hence, we can conclude that the energy storage at room temperature is viable only in ultra-thin AFE films ($h \leq 10$ nm) when the relative oxygen pressure does not deviate significantly from the normal conditions, $10^{-3} \leq \rho \leq 1$. Also, the energy storage is favorable in thicker films ($50 \leq h \leq 100$ nm) at elevated temperatures slightly above $T_A$ and relative pressures $\rho \leq 10^{-4}$.

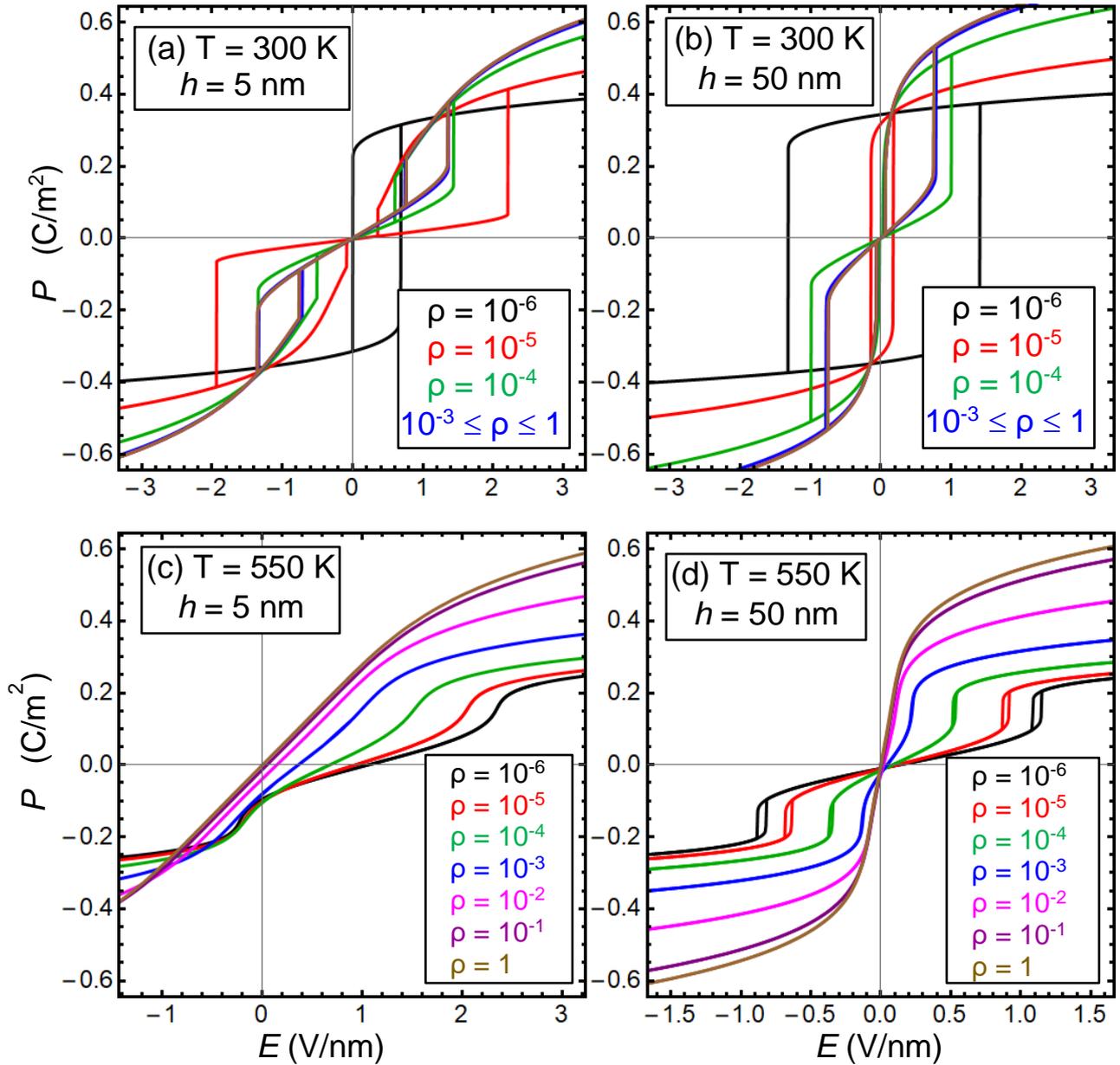

**FIGURE 7.** Polarization field dependence $P(E)$ calculated for a very low dimensionless frequency $w = 10^{-3}$, temperature $T = 300$ K (**a, b**) and 550 K (**c, d**), relative partial oxygen pressure $\rho = 10^{-6}$, (black curves), $10^{-5}$ (red curves), $10^{-4}$ (green curves), $10^{-3}$ (blue curves), $10^{-2}$ (magenta curves), $10^{-1}$ (purple curves) and 1 (brown

curves). The film thickness $h = 5$ nm (**a, b**), and 50 nm (**c, d**). External electric field is $E = \frac{U_0}{h}\sin(\omega t)$, and $w = \frac{\omega \Gamma_P}{2|\alpha_p|}$. Other parameters are the same as in **Fig. 3.**

To quantify the conclusion, we calculate the dependence of the hysteresis loop shape, LA and SA on the relative pressure, temperature and film thickness, and these results are shown in **Figs. E5-E7** in **Appendix E** [35]. After analyzing the results shown in **Fig. 7** and **Fig. E5-E7**, we compose a schematic diagram of the correlation between the loop shape, the energy loss, and the stored energy in dependence on the relative pressure $\rho$ and temperature $T$. It is shown in **Fig. 8a.** Color maps of the stored energy SA and loop area LA are shown in **Fig. 8b** and **8c**, respectively. The hysteresis loops map (black curves) is superimposed on the SA and LA color maps. As expected, the loop shape, and the SA and LA areas definitely correlate with the efficiency of the energy and information storage in the AFE films exposed to oxygen pressure.

From the diagram we conclude that the region I of double loops with a pronounced linear region is the most suitable for the high-density energy storage, and less suitable for the volatile information storage, because LA is high and SA is relatively small for them. The nonvolatile information storage is impossible for this type of double loops, since the spontaneous polarization $P_S$ is absent at zero and small voltages, $P_S(0) = 0$. The region I is the biggest: it corresponds to the pressures $10^{-3} \leq \rho \leq 1$ and temperatures $T < T_A$, and the area of the region slowly decreases with $\rho$ decrease or/and $T$ increase.

A region II corresponds to double loops without a linear region. The loops are characterized by high losses and thus are neither suitable for nonvolatile energy storage (since $P_S(0) = 0$ for the loops), nor for the effective energy storage (since their area LA is small). However, the type of loops is ideally suitable for a resistive-type (i.e., volatile) information storage, since SA is high for each minor loop. The region II has a shape of a curved stripe. The region corresponds to the pressures $10^{-3} \leq \rho \leq 1$ and temperatures $T < T_A$, its area increases with $\rho$ decrease; and it borders with the above region I.

The smallest region III contains single loops with a pronounced constriction, which are suitable for the information storage that is mostly nonvolatile, but the loops are less suitable for the energy storage, since LA is rather high and SA is very small. The boundaries of this region are diffuse, and it is located at pressures $10^{-6} \leq \rho \leq 10^{-3}$ and wide temperature range $T < 560$ K.

A region IV contains FE-like single AFI loops without a pronounced constriction, which are suitable for the nonvolatile information storage, but non-suitable for the energy storage. The region is a bit wider than region of AFI phase shown in **Fig. 6.** A region V, that is mostly PE phase, is suitable for the low-density energy storage in a definite sense, because the height of the polarization step is relatively small, but the loop area is very small (see green, black and red curves in **Fig. 7d**).

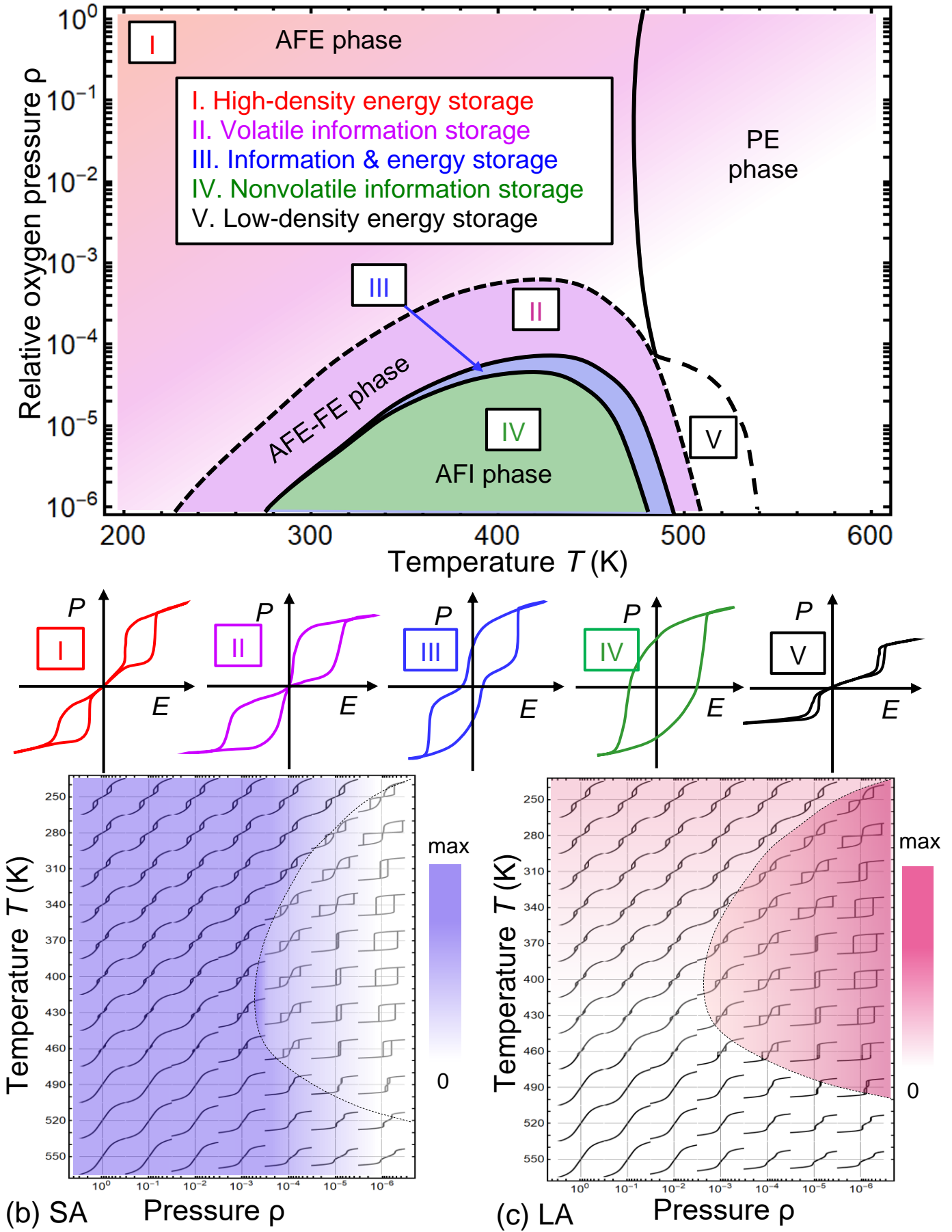

**FIGURE 8.** (a) A schematic diagram relating the loop shape, information and energy storage in dependence on the relative pressure $\rho$ and temperature $T$. Roman letters I - V are for the five regions described in the text. Color maps

of the stored energy SA **(b)** and loop area LA **(c)** of the $P(E)$ dependences. The hysteresis loops (black curves) are superimposed on the color maps. The film thickness $h = 5$ nm. Other parameters are the same as in **Fig. 3.**

## V. DISCUSSION AND OUTLOOK

We calculate the polar properties and phase diagram of a bulk antiferroelectric (prototype of PbZrO$_3$) using the modified 2-4-6 KLGD thermodynamic approach of two polarization sublattices for the description of the long-range polar and antipolar orderings.

Using the phenomenological parameters of KLGD thermodynamic potential, we explore the role of the surface ion layer with a charge density proportional to the partial oxygen pressure on the dipole states and their reversal mechanisms, and corresponding phase diagrams of AFE thin films using a Stephenson-Highland approach. The combined KLGD-SH approach allows to delineate the boundaries of the AFE, FE-like AFI, and electret-like PE states as a function of temperature, oxygen pressure, surface ions formation energy and concentration, and film thickness. This approach also allows the characterization of the polarization and anti-polarization dependence on the voltage applied to the antiferroelectric film, and the analysis of the static and dynamic hysteresis loop features. Important, that KLGD-SH approach proposes an alternative model for the frequently observed ferroelectricity in AFE thin films [51, 52] and provides a numerical model for the energy storage in AFI materials.

For applications, our modeling is able to select parameters, which can tune the position, where a transition to FE-like AFI state happens, minimize (or maximize) the area of hysteresis loops. Also, our results can be interesting for the implementation a multi-bit nonvolatile random-access memory (**NRAM**). As a matter of fact, a "single" hysteresis loop $P(E)$ with two values of the spontaneous polarization, $\pm P_S$; implements a "binary" bit in uniaxial ferroelectrics. In the case of a thin film of multiaxial ferroelectric, the coexistence of several phases with different directions of the polarization vector and its magnitude is possible. Recently, Baudry et al. [53] predict the existence of a multi-well free energy relief in a thin strained PbTiO$_3$ film, with the possibility of transitions between the wells under the action of an electric field. Depending on the misfit strain and temperature, 2 (c-phase), 3 (aa-phase) and 4 (r-phase) stable phases with different spontaneous polarizations are possible, which implements 4-bit, 3-bit or "normal" 2-bit. However, the transition between the multi-bits is possible only by changing the misfit strain, that is difficult to implement during the film exploitation in a memory cell. The considered AFI system, "antiferroelectric film + surface ions layer", allows to switching between 2 or 3 stable polarization states (see 2 or 3 yellow rectangles on the static loops in **Fig. 4**), which can implement ternary bits, and the transition of the nonvolatile memory cell to 3-bits is possible under the change of partial oxygen pressure, which can be easier from technological point of view and possible during the cell operation.

However, in this work we did not consider phases with several orientations of polarization and anti-polarization, leaving this for future studies, realizing that the multiaxiality can lead to the appearance of additional stable phases of the free energy and increase the number of multi-bits in the AFI-based NRAM.

To resume, our approach allows performing the overview of the phase diagrams of thin AFE films covered by surface ion layer and exploring the specifics of polarization reversal and antipolar ordering in the system, quantify the films applications for the energy and information storage, such as AFI NRAM. On the other hand, many important questions, such as the polarization multiaxiality, finite size effect and its influence on domain formation and evolution, remain for further studies.

**Acknowledgements** This effort is based upon work supported by the U.S. Department of Energy, Office of Science, Office of Basic Energy Sciences Energy Frontier Research Centers program under Award Number DE-SC0021118 (S.V.K) and performed at the Oak Ridge National Laboratory's Center for Nanophase Materials Sciences (CNMS), a U.S. Department of Energy, Office of Science User Facility. A.N.M. and N.V.M. work is supported by the National Academy of Sciences of Ukraine. This work was supported in part (A.B.) by the US Department of Energy, Office of Science, Office of Basic Energy Sciences, as part of the Energy Frontier Research Centers program: CSSAS–The Center for the Science of Synthesis Across Scales–under Award Number DE-SC0019288, located at University of Washington and performed at Oak Ridge National Laboratory's Center for Nanophase Materials Sciences (CNMS), a U.S. Department of Energy, Office of Science User Facility. A.N.M work is supported by the National Research Foundation of Ukraine (Grant application Φ81/41481).

**Authors' contribution.** S.V.K. and A.N.M. generated the research idea. A.N.M. proposed the theoretical model, derived analytical results and interpreted numerical results, obtained by E.A.E. A.B. used Gaussian Process model for rapid exploration and prediction of phase diagrams. A.N.M. and S.V.K. wrote the manuscript draft. Then S.V.K. and N.V.M. worked on the results discussion and manuscript improvement.

# REFERENCES

[1] A.K. Tagantsev, L.E. Cross, and J. Fousek, *Domains in ferroic crystals and thin films* (Springer, New York, 2010).

[2] A.M. Bratkovsky, and A.P. Levanyuk. Continuous theory of ferroelectric states in ultrathin films with real electrodes. J. Comput. Theor. Nanoscience **6**, 465 (2009).

[3] A.M. Bratkovsky, and A.P. Levanyuk. Effects of anisotropic elasticity in the problem of domain formation and stability of monodomain state in ferroelectric films. Phys. Rev. B **84**, 045401 (2011).

[4] S.V. Kalinin and D.A. Bonnell, Screening Phenomena on Oxide Surfaces and Its Implications for Local Electrostatic and Transport Measurements. Nano Lett. **4**, 555 (2004).

[5] S. Jesse, A.P. Baddorf, and S.V. Kalinin, Switching spectroscopy piezoresponse force microscopy of ferroelectric materials. Appl. Phys. Lett. **88**, 062908 (2006).

[6] A.N. Morozovska, S.V. Svechnikov, E.A. Eliseev, S. Jesse, B.J. Rodriguez, and S.V. Kalinin, Piezoresponse force spectroscopy of ferroelectric-semiconductor materials. J. Appl. Phys. **102**, 114108 (2007).

[7] A.V. Ievlev, S. Jesse, A.N. Morozovska, E. Strelcov, E.A. Eliseev, Y.V. Pershin, A. Kumar, V.Y. Shur, and S.V. Kalinin, Intermittency, quasiperiodicity and chaos in probe-induced ferroelectric domain switching. Nat. Phys. **10**, 59 (2013).

[8] A.V. Ievlev, A.N. Morozovska, V.Ya. Shur, S.V. Kalinin. Humidity effects on tip-induced polarization switching in lithium niobate. Appl. Phys. Lett. **104**, 092908 (2014).

[9] E.V. Chensky and V.V. Tarasenko, Theory of phase transitions to inhomogeneous states in finite ferroelectrics in an external electric field. Sov. Phys. JETP **56**, 618 (1982) [Zh. Eksp. Teor. Fiz. **83**, 1089 (1982)].

[10] M. Stengel, and N.A. Spaldin, Origin of the dielectric dead layer in nanoscale capacitors, Nature 443, 679 (2006).

[11] N.Domingo, I. Gaponenko, K. Cordero-Edwards, Nicolas Stucki, V. Pérez-Dieste, C. Escudero, E. Pach, A. Verdaguer, and P. Paruch. Surface charged species and electrochemistry of ferroelectric thin films. Nanoscale **11**, 17920 (2019).

[12] J.E. Spanier, A.M. Kolpak, J.J. Urban, I. Grinberg, L. Ouyang, W.S.Yun, A.M. Rappe, and Hongkun Park. Ferroelectric phase transition in individual single-crystalline $BaTiO_3$ nanowires. Nano Lett. **6**, 735 (2006).

[13] R.V. Wang, D.D. Fong, F. Jiang, M.J. Highland, P.H. Fuoss, C. Tompson, A.M. Kolpak, J.A. Eastman, S.K. Streiffer, A.M. Rappe, and G.B. Stephenson, Reversible chemical switching of a ferroelectric film, Phys. Rev. Lett. **102**, 047601 (2009).

[14] D.D. Fong, A.M. Kolpak, J.A. Eastman, S.K. Streiffer, P.H. Fuoss, G.B. Stephenson, C.Thompson, D.M. Kim, K.J. Choi, C.B. Eom, I. Grinberg, and A.M. Rappe. Stabilization of Monodomain Polarization in Ultrathin $PbTiO_3$ Films. Phys. Rev. Lett. **96**, 127601 (2006).

[15] M.J. Highland, T.T. Fister, M.-I. Richard, D.D. Fong, P.H. Fuoss, C.Thompson, J.A. Eastman, S.K. Streiffer, and G.B.Stephenson. Polarization Switching without Domain Formation at the Intrinsic Coercive Field in Ultrathin Ferroelectric $PbTiO_3$. Phys. Rev. Lett. **105**, 167601 (2010).

[16] J.D. Baniecki, J.S. Cross, M. Tsukada, and J. Watanabe. H$_2$O vapor-induced leakage degradation of Pb(Zr,Ti)O$_3$ thin-film capacitors with Pt and IrO$_2$ electrodes. Appl. Phys. Lett. **81**, 3837 (2002).

[17] Y. Gu, K. Xu, C. Song, X. Zhong, H. Zhang, H. Mao, M.S. Saleem, J. Sun, W. Liu, Z. Zhang, F. Pan, and J. Zhu, Oxygen-Valve Formed in Cobaltite-Based Heterostructures by Ionic Liquid and Ferroelectric Dual-Gating. ACS Appl. Mater. Interfaces **11**, 19584 (2019).

[18] N.C. Bristowe, M. Stengel, P.B. Littlewood, J.M. Pruneda, and E. Artacho. Electrochemical ferroelectric switching: Origin of polarization reversal in ultrathin films. Phys. Rev. B **85**, 024106 (2012).

[19] S.V. Kalinin, and D.A. Bonnell, Effect of phase transition on the surface potential of the BaTiO3 (100) surface by variable temperature scanning surface potential microscopy. J. Appl. Phys. **87**, 3950 (2000).

[20] S.V. Kalinin, C.Y. Johnson, D.A. Bonnell, Domain polarity and temperature induced potential inversion on the BaTiO$_3$(100) surface. J. Appl. Phys. **91**, 3816 (2002).

[21] S. Buhlmann, E. Colla, P. Muralt, Polarization reversal due to charge injection in ferroelectric films. Phys. Rev. B **72**, 214120 (2005).

[22] Y. Kim, S. Buhlmann, S. Hong, S.H. Kim, K. No, Injection charge assisted polarization reversal in ferroelectric thin films. Appl. Phys. Lett. **90**, 072910 (2007).

[23] Y. Kim, J. Kim, S. Bühlmann, S. Hong, Y.K. Kim, S.-H. Kim, and K. No. Screen charge transfer by grounded tip on ferroelectric surfaces. Phys. Status Solidi (RRL) **2**, 74 (2008).

[24] A.V. Ievlev, A.N. Morozovska, E.A. Eliseev, V.Y. Shur, and S.V. Kalinin. Ionic field effect and memristive phenomena in single-point ferroelectric domain switching. Nat. Commun. **5**, art. no. 4545 (2014).

[25] G.B. Stephenson and M.J. Highland, Equilibrium and stability of polarization in ultrathin ferroelectric films with ionic surface compensation. Phys. Rev. B, **84**, 064107 (2011).

[26] H. Lu, C-W. Bark, D.E. De Los Ojos, J. Alcala, C.-B. Eom, G. Catalan, and A. Gruverman. Mechanical writing of ferroelectric polarization. Science **336**, 59 (2012).

[27] C.W. Bark, P. Sharma, Y. Wang, S.H. Baek, S. Lee, S. Ryu, C.M. Folkman, T.R. Paudel, A. Kumar, S.V. Kalinin, A. Sokolov, E.Y. Tsymbal, M.S. Rzchowski, A. Gruverman, C.B. Eom, Switchable induced polarization in LaAlO$_3$/SrTiO$_3$ heterostructures. Nano Lett. **12**, 1765 (2012).

[28] Y. Cao, A.N. Morozovska, and S.V. Kalinin. Pressure-induced switching in ferroelectrics: Phase-field modeling, electrochemistry, flexoelectric effect, and bulk vacancy dynamics. Phys. Rev. B **96**, 184109, (2017).

[29] M.J. Highland, T.T. Fister, D.D. Fong, P.H. Fuoss, C. Thompson, J.A. Eastman, S.K. Streiffer, and G.B. Stephenson. Equilibrium polarization of ultrathin PbTiO3 with surface compensation controlled by oxygen partial pressure. Phys. Rev. Lett. **107**, 187602 (2011).

[30] S.M. Yang, A.N. Morozovska, R. Kumar, E.A. Eliseev, Y. Cao, L. Mazet, N. Balke, S. Jesse, R. Vasudevan, C. Dubourdieu, S.V. Kalinin. Mixed electrochemical-ferroelectric states in nanoscale ferroelectrics. Nat. Phys. **13**, 812 (2017).

[31] A.N. Morozovska, E.A. Eliseev, N.V. Morozovsky, and S.V. Kalinin. Ferroionic states in ferroelectric thin films. Phys. Rev. B **95**, 195413 (2017).

[32] A.N. Morozovska, E.A. Eliseev, N.V. Morozovsky, and S.V. Kalinin. Piezoresponse of ferroelectric films in ferroionic states: time and voltage dynamics. Appl. Phys. Lett. **110**, 182907 (2017).

[33] S.V. Kalinin, Y. Kim, D.D. Fong, and A.N. Morozovska. Surface Screening Mechanisms in Ferroelectric Thin Films and its Effect on Polarization Dynamics and Domain Structures. Rep. Prog. Phys. **81**, 036502 (2018).

[34] A.N. Morozovska, E.A. Eliseev, A.I. Kurchak, N.V. Morozovsky, R.K. Vasudevan, M.V. Strikha, and S.V. Kalinin. Effect of surface ionic screening on polarization reversal scenario in ferroelectric thin films: crossover from ferroionic to antiferroionic states. Phys. Rev. B **96**, 245405 (2017).

[35] See Supplementary Materials for details [URL will be provided by Publisher]

[36] A.K. Tagantsev and G. Gerra. Interface-induced phenomena in polarization response of ferroelectric thin films. J. Appl. Phys. **100**, 051607 (2006).

[37] R. Blinc and B. Zeks, Soft Mode in Ferroelectrics and Antiferroelectrics (North-Holland Publishing Company, Amsterdam, Oxford, 1974).

[38] R. Kretschmer, K. Binder, Surface effects on phase transitions in ferroelectrics and dipolar magnets. Phys. Rev. B **20**, 1065 (1979).

[39] C.-L. Jia, V. Nagarajan, Jia-Q.He, L. Houben, T.Zhao, Ramamoorthy Ramesh, K.Urban, and R. Waser. Unit-cell scale mapping of ferroelectricity and tetragonality in epitaxial ultrathin ferroelectric films. Nat. Mat. **6**, 64 (2007).

[40] K.Y. Foo, and B.H. Hameed. Insights into the modeling of adsorption isotherm systems. Chem. Eng. J. **156,** 2 (2010).

[41] A.I. Kurchak, A.N. Morozovska, S.V. Kalinin, and M.V. Strikha. Nontrivial temperature behavior of the carriers concentration in the nano-structure "graphene channel - ferroelectric substrate with domain walls. Acta Mater. **155**, 302 (2018).

[42] Y. Cao, and S.V. Kalinin. Phase-field modeling of chemical control of polarization stability and switching dynamics in ferroelectric thin films. Phys. Rev. B **94**, 235444 (2016).

[43] B.D. Vujanovic, S.E. Jones, Variational Methods in Nonconservative Phenomena (Academic Press, San Diego, 1989)

[44] M.J. Haun, T.J. Harvin, M.T. Lanagan, Z.Q. Zhuang, S.J. Jang, and L.E. Cross, Thermodynamic theory of $PbZrO_3$, J. Appl. Phys. **65**, 3173 (1989).

[45] E. Savaguchi, T. Kittaka. Antiferroelectricity and ferroelectricity in lead circonate. J. Phys. Soc. Japan **7**, 336 (1952)

[46] E. Savaguchi. Ferroelectricity versus antiferroelectricity in the solid solutions of $PbZrO_3$ and $PbTiO_3$. J. Phys. Soc. Japan, **8**, 615 (1953)

[47] R.H. Dungan, H.M. Barnett, A.H. Stark. Phase relations and electrical parameters in the ferroelectric-antiferroelectric region of the system $PbZrO_3$-$PbTiO_3$-$PbNbO_2O_6$". J. Am. Ceram. Soc. **45**, 382 (1962).

[48] P. Dufour, T. Maroutian, A. Chanthbouala, C. Jacquemont, F. Godel, L. Yedra, M. Otonicar, N. Guiblin, M. Bibes, B. Dkhil, S. Fusil and V. Garcia. Electric-field and temperature induced phase transitions in antiferroelectric

thin films of PbZrO$_3$. ISAF 2021,

https://epapers.org/isaf2021/ESR/paper_details.php?PHPSESSID=fu0r61aer8lkagu8v4ncivu7o4&paper_id=3520

[49] A. S. Mischenko, Qi Zhang, J. F. Scott, R. W. Whatmore, and N. D. Mathur. Giant electrocaloric effect in thin-film PbZr$_{0.95}$Ti$_{0.05}$O$_3$. Science **311**, no. 5765, 1270 (2006)

[50] A.K. Tagantsev, K. Vaideeswaran, S.B. Vakhrushev, A.V. Filimonov, R.G. Burkovsky, A. Shaganov, D. Andronikova, A.I. Rudskoy, A.Q.R. Baron, H. Uchiyama, D. Chernyshov, A. Bosak, Z. Ujma, K. Roleder, A. Majchrowski, J.-H. Ko, and N. Setter, The origin of antiferroelectricity in PbZrO$_3$. Nat. Comm. 4, art. no. 2229 (2013).

[51] S. Chattopadhyay, P. Ayyub, V.R. Palkar, M.S. Multani, S.P. Pai, S.C. Pai, S.C. Purandare and R. Pinto, J. Appl. Phys., **83**, 7808 (1998).

[52] S. Chattopadhyay, P. Ayyub, V.R. Palkar, M.S. Multani, S.P. Pai, S.C. Pai, S.C. Purandare and R. Pinto, J. Appl. Phys., **83**, 7808 (1998).

[53] L. Baudry, I. Lukyanchuk, and V.M. Vinokur, Ferroelectric symmetry-protected multibit memory cell. Scientific Reports **7**, 42196 (2017).

**SUPPLEMETARY MATERIALS to**

**"Effect of surface ionic screening on polarization reversal and phase diagrams in thin antiferroelectric films for information and energy storage"**

## APPENDIX A.
### Electrostatic equations with boundary conditions

Quasi-static electric field inside the ferroelectric film is defined via electric potential $\phi_f$ in a conventional way, $E_3 = -\partial \phi_f/\partial x_3$. The potential $\phi_f$ satisfies electrostatic equations in the gap and in the ferroelectric film, which acquires the form:

$$\Delta \phi_d = 0, \qquad \text{(inside the gap } -\lambda \leq z \leq 0) \qquad (A.1a)$$

$$\left(\varepsilon_{33}^b \frac{\partial^2}{\partial z^2} + \varepsilon_{11}^f \Delta_\perp\right)\phi_f = \frac{1}{\varepsilon_0}\frac{\partial P_3^f}{\partial z}, \quad \text{(inside the ferroelectric film } 0 < z < h) \quad (A.1b)$$

where $\Delta$ is 3D-Laplace operator is $\Delta$, $\Delta_\perp$ is 2D-Laplace operator.

Boundary conditions to the system (A.1) have the form:

$$\phi_d|_{z=-\lambda} = U, \quad (\phi_d - \phi_f)|_{z=0} = 0, \quad \phi_f|_{z=h} = 0, \qquad (A.2a)$$

$$\left(\varepsilon_0 \varepsilon_d \frac{\partial \phi_d}{\partial z} + P_3^f - \varepsilon_0 \varepsilon_{33}^b \frac{\partial \phi_f}{\partial z} - \sigma\right)\bigg|_{z=0} = 0. \qquad (A.2b)$$

We use the 2-4-6 Kittel-type models for equivalent dipole sublattices. Corresponding Kittel energy, written in $P_i^{(j)}$ representation, is

$$G_{Kittel} = \frac{\alpha}{2}\left[\left(P_3^{(1)}\right)^2 + \left(P_3^{(2)}\right)^2\right] + \beta P_3^{(1)} P_3^{(2)} + \frac{\gamma}{4}\left(P_3^{(1)}\right)^2\left(P_3^{(2)}\right)^2 + \frac{\delta}{4}\left[\left(P_3^{(1)}\right)^4 + \left(P_3^{(2)}\right)^4\right] +$$

$$\frac{g}{2}\left[\left(\frac{\partial P_3^{(1)}}{\partial z}\right)^2 + \left(\frac{\partial P_3^{(1)}}{\partial z}\right)^2\right] - \left[P_3^{(1)} + P_3^{(2)}\right]\frac{E_3}{2} \quad (A.3)$$

Making here the substitution $P_i = \frac{1}{2}\left(P_i^{(1)} + P_i^{(2)}\right)$ and $A_i = \frac{1}{2}\left(P_i^{(1)} - P_i^{(2)}\right)$, we lead to the expression

$$G_{Kittel} = (\alpha+\beta)P_3^2 + (\alpha-\beta)A_3^2 + \left(3\delta - \frac{\gamma}{2}\right)P_3^2 A_3^2 + \left(\frac{\gamma}{4}+\frac{\delta}{2}\right)(P_3^4+A_3^4) - P_3 E_3 + \frac{g}{2}\left[\left(\frac{\partial P_3}{\partial z}\right)^2 + \left(\frac{\partial A_3}{\partial z}\right)^2\right]$$

(A.4a)

Modification of Eqs.(A.3) and (A.4) allowing for the six-order invariants, $\frac{\mu}{4}\left[\left(P_3^{(1)}\right)^4\left(P_3^{(2)}\right)^2 + \left(P_3^{(1)}\right)^2\left(P_3^{(2)}\right)^4\right] + \epsilon\left(P_3^{(1)}\right)^3\left(P_3^{(2)}\right)^3$, and oxygen tilt leads to the free energy, which formally coincides with the form used by Haun et al (see Appendix C):

$$G_{Kittel} = \alpha_p P_3^2 + \alpha_a A^2 + \chi P_3^2 A^2 + \beta_p P_3^4 + \beta_a A^4 + \gamma_p P_3^6 + \gamma_a A^6 - P_3 E_3 \qquad (A.4b)$$

# APPENDIX B.

## Free energy with renormalized coefficients

Below we assume that the polarization distribution $P_3(x, y, z)$ is smooth enough, the coupled nonlinear algebraic equations for the polarization $P$ averaged over film thickness and surface charge density $\sigma$ are valid:

$$\Gamma \frac{\partial P}{\partial t} + 2\alpha_p P_3 + 4\beta_p P_3^3 + 2\chi P_3 A^2 + 6\gamma_p P^5 = \frac{\Psi(U,\sigma,P)}{h}, \qquad (B.1a)$$

$$\Gamma \frac{\partial A}{\partial t} + 2\alpha_a A + 4\beta_a A^3 + 2\chi P^2 A + 6\gamma_a A^5 = 0, \qquad (B.1b)$$

$$\tau \frac{\partial \sigma}{\partial t} + \sigma = \sigma_0[\Psi(U, \sigma, P)]. \qquad (B.1c)$$

The overpotential is given by expression $\Psi(U, \sigma, P) = \frac{\lambda(\sigma-P)+\varepsilon_0 \varepsilon_d U}{\varepsilon_0(\varepsilon_d h + \lambda \varepsilon_{33}^b)} h$ and the function $\sigma_0[\psi] = \sum_i \frac{e Z_i \theta_i(\psi)}{A_i} \equiv \sum_i \frac{e Z_i}{A_i} \left(1 + \rho^{1/n_i} exp\left(\frac{\Delta G_i^{00} + e Z_i \psi}{k_B T}\right)\right)^{-1}$. Electric potentials acting in the dielectric gap ($\phi_d$) and in the ferroelectric film ($\phi_f$) linearly depends on the coordinate z and overpotential, namely $\phi_d = U - \frac{z+\lambda}{\lambda}(U - \Psi)$ and $\phi_f = (h - z)\frac{\Psi}{h}$. Below we'll use the designations

$$\Delta G_i^{0p} = \Delta G_i^{00} + \frac{k_B T}{n_i} \ln(\rho). \qquad (B.1d)$$

Next we consider the stationary case, when one can put $\sigma = \sigma_0[\Psi(U, \sigma, P)]$ in Eqs.(B.1). Corresponding free energy $G[P, \Psi]$ that's formal minimization, $\frac{\partial G[P,\Psi]}{\partial P} = 0$ and $\frac{\partial G[P,\Psi]}{\partial \Psi} = 0$, leads to Eqs.(B.1), has the form:

$$\frac{G[P,\Psi]}{S} = h(\alpha_p P^2 + \alpha_a A^2 + \chi P^2 A^2 + \beta_p P^4 + \beta_a A^4 + \gamma_p P^6 + \gamma_a A^6) - \Psi P - \varepsilon_0 \varepsilon_{33}^b \frac{\Psi^2}{2h} - \frac{\varepsilon_0 \varepsilon_d}{2} \frac{(\Psi - U)^2}{\lambda} + \int_0^\Psi \sigma_0[\varphi] d\varphi, \qquad (B.2a)$$

Here $\alpha_p = \alpha_{pT}(T - T_C) + \frac{2g}{h\Lambda_P}$ and $\alpha_a = \alpha_{aT}(T - \Theta) + \frac{2g}{h\Lambda_A}$. For the sake of simplicity, we may regard that either $\Lambda_A \to \infty$ or $\Lambda_P = \Lambda_A = \Lambda$.

The energy (B.2a) has absolute minima at high $\Psi$ values. So, according to the Biot's variational principle, let us find for the incomplete thermodynamic potential, which partial minimization over $P$ will give the equations of state, and, at the same time, it has an absolute minimum at finite $P$ values. Substituting here the expression for the overpotential $\frac{\Psi}{h} = \frac{\lambda(\sigma_0[\Psi]-P)+\varepsilon_0 \varepsilon_d U}{\varepsilon_0(\varepsilon_d h + \lambda \varepsilon_{33}^b)}$ we derived the single equation for the average polarization:

$$2\alpha_p P + 4\beta_p P^3 + 2\chi P A^2 + 6\gamma_p P^5 = \frac{\lambda}{\varepsilon_0(\varepsilon_d h + \lambda \varepsilon_{33}^b)} \sum_{i=1,2} \frac{e Z_i}{A_i} \left(1 + exp\left[\frac{\Delta G_i^{0p} + e Z_i h(2\alpha_p P + 4\beta_p P^3 + 2\chi P A^2 + 6\gamma_p P^5)}{k_B T}\right]\right)^{-1} - \frac{\lambda P - \varepsilon_0 \varepsilon_d U}{\varepsilon_0(\varepsilon_d h + \lambda \varepsilon_{33}^b)} \qquad (B.3)$$

Corresponding potential, which minimization over $P$ gives Eq.(B.3), has the form:

$$F[P,A] = \left(\frac{\lambda}{2\varepsilon_0(\varepsilon_d h + \lambda\varepsilon_{33}^b)} + \alpha_p\right)P^2 + \alpha_a A^2 + \chi P^2 A^2 + \beta_p P^4 + \beta_a A^4 + \gamma_p P^6 + \gamma_a A^6 + \frac{\varepsilon_d UP}{\varepsilon_d h + \lambda\varepsilon_{33}^b} -$$

$$-\frac{\lambda}{\varepsilon_0(\varepsilon_d h + \lambda\varepsilon_{33}^b)}\sum_{i=1,2}\frac{eZ_i}{A_i}\int_0^P dp\left(1 + exp\left(\frac{\Delta G_i^{0p} + eZ_i h(2\alpha_p p + 4\beta_p p^3 + 2\chi p A^2 + 6\gamma_p p^5)}{k_B T}\right)\right)^{-1}$$

(B.4)

Next, we consider the condition $\left|\frac{eZ_i \Psi}{k_B T}\right| \ll 1$ and use that

$$\left(1 + exp\left(\frac{\Delta G_i^{0p} + eZ_i \Psi}{k_B T}\right)\right)^{-1} \approx f_i(T,\rho)\left(1 - \frac{eZ_i \Psi}{k_B T}f_i(T,\rho)\right) \quad \text{(B.5)}$$

were we introduced the sort of "level filling factor" $f_i(T,\rho) = \left(1 + exp\left(\frac{\Delta G_i^{0p}}{k_B T}\right)\right)^{-1}$

In this case the expression (B.4) can be further simplified as

$$\left(\frac{\lambda}{2\varepsilon_0(\varepsilon_d h + \lambda\varepsilon_{33}^b)} + \alpha_p\right)P^2 + \alpha_a A^2 + \chi P^2 A^2 + \beta_p P^4 + \beta_a A^4 + \gamma_p P^6 + \gamma_a A^6 + \frac{\varepsilon_d UP}{\varepsilon_d h + \lambda\varepsilon_{33}^b}$$

$$-\frac{\lambda}{\varepsilon_0(\varepsilon_d h + \lambda\varepsilon_{33}^b)}\sum_{i=1,2}\frac{eZ_i}{A_i}f_i(T,\rho)\int_0^P dp\left(1 - f_i(T,\rho)\frac{eZ_i h}{k_B T}(2\alpha_p p + 4\beta_p p^3 + 2\chi p A^2 + 6\gamma_p p^5)\right)$$

(B.6)

Using that

$$\int_0^P dp\left(1 - f_i(T,\rho)\frac{eZ_i h}{k_B T}(2\alpha_p p + 4\beta_p p^3 + 2\chi p A^2 + 6\gamma_p p^5)\right) = P - f_i(T,\rho)\frac{eZ_i h}{k_B T}\left(\alpha_p P^2 + \beta_p P^4 + \chi P^2 A^2 + \gamma_p P^6\right) \quad \text{(B.7)}$$

One could get from (B.6):

$$F[P,A] = \left(\alpha_p(1 + S(T,\rho,h)) + \frac{\lambda}{2\varepsilon_0(\varepsilon_d h + \lambda\varepsilon_{33}^b)}\right)P^2 + (1 + S(T,\rho,h))\left(\chi P^2 A^2 + \beta_p P^4 + \gamma_p P^6\right) -$$

$$\left(\frac{\lambda}{\varepsilon_0(\varepsilon_d h + \lambda\varepsilon_{33}^b)}\sum_{i=1,2}\frac{eZ_i}{A_i}f_i(T,\rho) - \frac{\varepsilon_d U}{\varepsilon_d h + \lambda\varepsilon_{33}^b}\right)P + \alpha_a A^2 + \beta_a A^4 + \gamma_a A^6 \quad \text{(B.7a)}$$

The function $S$ in Eqs.(B.5b) and (B.5c) is

$$S(T,\rho,h) = \frac{\lambda h}{\varepsilon_0(\varepsilon_d h + \lambda\varepsilon_{33}^b)}\sum_{i=1,2}\frac{(eZ_i f_i(T,\rho))^2}{k_B T A_i} \quad \text{(B.7b)}$$

The effective field, produced by ionic charge and applied potential $U$, has the form:

$$E_{eff}(U,\Delta G_i^{00}) = \frac{\lambda}{\varepsilon_0(\varepsilon_d h + \lambda\varepsilon_{33}^b)}\sum_{i=1,2}\frac{eZ_i}{A_i}f_i(T,\rho) - \frac{\varepsilon_d U}{\varepsilon_d h + \lambda\varepsilon_{33}^b}, \quad \text{(B.8)}$$

## APPENDIX C

**Estimation of LGD expansion coefficients for PbZrO3 from the papers of Haun et al.**

Bulk free energy expansion on the powers of polarization P and anti-polarization A components of vector has the following form in the case of m3m symmetry of paraelectric phase:

$$G_{Haun} = a_1(P_1^2 + P_2^2 + P_3^2) + a_{11}(P_1^4 + P_2^4 + P_3^4) + a_{12}(P_1^2 P_2^2 + P_1^2 P_3^2 + P_2^2 P_3^2) + a_{111}(P_1^6 + P_2^6 + P_3^6)$$
$$+ a_{112}(P_1^2(P_2^4 + P_3^4) + P_1^4(P_2^2 + P_3^2) + P_2^2 P_3^4 + P_2^4 P_3^2) + a_{123} P_1^2 P_2^2 P_3^2$$
$$+ \sigma_1(A_1^2 + A_2^2 + A_3^2) + \sigma_{11}(A_1^4 + A_2^4 + A_3^4) + \sigma_{12}(A_1^2 A_2^2 + A_1^2 A_3^2 + A_3^2 A_2^2)$$
$$+ \sigma_{111}(A_1^6 + A_2^6 + A_3^6) + \sigma_{112}(A_1^2(A_2^4 + A_3^4) + A_1^4(A_2^2 + A_3^2) + A_2^2 A_3^4 + A_2^4 A_3^2)$$
$$+ \sigma_{123} A_1^2 A_2^2 A_3^2 + \mu_{11}(P_1^2 A_1^2 + P_2^2 A_2^2 + P_3^2 A_3^2)$$
$$+ \mu_{12}(P_1^2 A_2^2 + P_2^2 A_1^2 + P_2^2 A_3^2 + P_3^2 A_2^2 + P_3^2 A_1^2 + P_1^2 A_3^2) + \mu_{44}(P_1 P_2 A_1 A_2 + P_2 P_3 A_2 A_3$$
$$+ P_1 P_3 A_1 A_3)$$

(C.1a)

Here two coefficients are supposed to be linearly dependent on the temperature

$$a_1 = \frac{T - T_P}{2\varepsilon_0 C_P}, \quad \sigma_1 = \frac{T - T_A}{2\varepsilon_0 C_A} \tag{C.1b}$$

It was pointed by Haun et al, that in accordance with experiment an orthorhombic antiferroelectric phase is stable in a wide temperature range (with $A_1 = \pm A_2 \neq 0$ and $A_3 = 0$), while a rhombohedral ferroelectric phase (with the three components of polarization equal, $P_1 = \pm P_2 = \pm P_3 \neq 0$) is possible in narrow temperature range up to 232°C. For these two phases free energy (C.1) could be simplified as follows

$$G_{Haun} = 3a_1 P_3^2 + 3(a_{11} + a_{12})P_3^4 + (3a_{111} + 6a_{112} + a_{123})P_3^6 + 2\sigma_1 A_1^2 + (2\sigma_{11} + \sigma_{12})A_1^4$$
$$+ (2\sigma_{111} + 2\sigma_{112})A_1^6 + (2\mu_{11} + 4\mu_{12} + \mu_{44})(P_1^2 A_1^2)$$
$$\equiv |P_1^2 + P_2^2 + P_3^2 = P^2 \equiv 3P_3^2 \text{ and } A_1^2 + A_2^2 = A^2 \equiv 2A_1^2|$$
$$\equiv a_1 P^2 + \frac{a_{11} + a_{12}}{3} P^4 + \frac{3a_{111} + 6a_{112} + a_{123}}{27} P^6 +$$
$$\sigma_1 A^2 + \frac{2\sigma_{11} + \sigma_{12}}{4} A^4 + \frac{\sigma_{111} + \sigma_{112}}{4} A^6 + \frac{2\mu_{11} + 4\mu_{12} + \mu_{44}}{6} P^2 A^2$$

(C.2)

Polarization "part" of the energy (C.2) could be extrapolated as a limiting case of PZT solid solution when the Ti content tends to zero (using coefficients taken from Haun et al. [1]). In this way we obtained that

$$T_P = 463\,K, C_P = 2.027\,10^5\,K, \beta_p \equiv \frac{a_{11} + a_{12}}{3} = -3.825 \times 10^8, \gamma_p \equiv \frac{3a_{111} + 6a_{112} + a_{123}}{27} = 3.12591 \times 10^9$$

(C.3)

The situation with "anti-polar part" of (C.2) is more complex, since the detailed information on the anti-ferroelectric phase of PZO is still not available. Haun et al. was able to estimate only some combinations of the parameters, for instance, only the product of corresponding Curie-Weiss constant, $C_A$, and the coefficient of "anti-electrostriction", $Z_{44}$, was found as $C_A Z_{44} = 683.35\,K\,m^4/C^2$. Under supposition that "anti-electrostriction" is the same as electrostriction, $Z_{44} \cong Q_{44} = 0.059$, we found that

$$T_A = 490\,K, \quad C_A = 1.16\,10^4\,K. \tag{C.4}$$

The rest of the necessary expansion coefficients could be found without additional assumptions. For instance, coupling coefficient was found from the relation

$$\lambda = 2\frac{\mu_{11}+2\mu_{12}}{3\,Z_{44}} = 6.0184\,10^{11}\frac{m}{F} \Rightarrow \chi \sim \frac{\mu_{11}+2\mu_{12}}{3} = 1.775428 \times 10^{10} \quad (C.5)$$

Recalling spontaneous shear strain $u_4 = Z_{44}A_1^2 \cong 0.00133$ we estimated corresponding value of $A_1 \cong 0.150\,C/m^2$. Finally, using the relations between different coefficients at Neel temperature ($T_N=226°C$)

$$(2\sigma_1 + (2\sigma_{11} + \sigma_{12})A_1^2 + 2(\sigma_{111} + \sigma_{112})A_1^4)|_{T=T_N} = 0$$

$$2\sigma_1 + 2(2\sigma_{11} + \sigma_{12})A_1^2 + 6(\sigma_{111} + \sigma_{112})A_1^4 = 0$$

one could easily find the values

$$\beta_a = \frac{2\sigma_{11}+\sigma_{12}}{4} \approx -2.0561 \times 10^9 \text{ and } \gamma_a = \frac{\sigma_{111}+\sigma_{112}}{4} \approx 2.2803 \times 10^{10} \quad (C.6)$$

**TABLE C1. Description of physical variables and their numerical values**

| Description of main physical quantities used in Eqs.(1)-(3) | Designation and dimensionality | Value for a structure PbZrO$_3$ film / ionic charge / gap / tip |
|---|---|---|
| Polarization of ferroelectric along polar axis Z | $P_3$ (C/m$^2$) | variable (0.26 for a bulk material) |
| Electric field | $E_3$ (V/m) | variable |
| Electrostatic potentials of dielectric gap and ferroelectric film | $\varphi_d$ and $\varphi_f$ (V) | variables |
| Electric voltage on the tip | $U$ (V) | variable |
| Coefficient of LGD functional | $\alpha_p = \alpha_{pT}(T - T_C)$ (C$^{-2}$ J m) | T-dependent variable |
| Dielectric stiffness | $\alpha_{pT}$ (×10$^5$ C$^{-2}$·J·m/K) | 2.7969 |
| Curie temperature | $T_P$ (K) | 463.2 |
| Coefficient of LGD functional | $\alpha_a = \alpha_{aT}(T - \Theta)$ (C$^{-2}$ J m) | T-dependent variable |
| | $\alpha_{aT}$ (×10$^6$ C$^{-2}$·J·m/K) | 4.8789 |
| AFE temperature | $T_A$ (K) | 490 |
| Coefficient of LGD functional | $\beta_p$ (×10$^8$ J C$^{-6}$·m$^9$) | -3.825 |
| Coefficient of LGD functional | $\beta_a$ (×10$^9$ J C$^{-6}$·m$^9$) | -2.056 |
| Coefficient of LGD functional | $\chi$ (×10$^{10}$ J C$^{-6}$·m$^9$) | 1.7754 |
| Coefficient of LGD functional | $\gamma_p$ (×10$^9$ J C$^{-8}$·m$^{13}$) | 3.126 |
| Coefficient of LGD functional | $\gamma_a$ (×10$^{10}$ J C$^{-8}$·m$^{13}$) | 2.280 |
| Gradient coefficient | $g$ (×10$^{-10}$ m/F) | (0.5-5) |
| Kinetic coefficient | $\Gamma$ (s× C$^{-2}$ J m) | rather small |
| Landau-Khalatnikov relaxation time | $\tau_K$ (s) | 10–11 – 10–13 (far from Tc) |
| Thickness of ferroelectric layer | $h$ (nm) | 50 (variable) |
| Background permittivity of ferroelectric | $\varepsilon_{33}^b$ (dimensionless) | 10 |
| Extrapolation lengths | $\Lambda-$, $\Lambda+$ (angstroms) | $\Lambda_-=1$ Å, $\Lambda_+=2$ Å |
| Surface charge density | $\sigma(\phi,t)$ (C/m$^2$) | variable |
| Equilibrium surface charge density | $\sigma_0(\phi)$ (C/m$^2$) | variable |
| Occupation degree of surface ions | $\theta i$ (dimensionless) | variable |
| Relative oxygen partial pressure | $\rho$ (dimensionless) | variable |
| Surface charge relaxation time | $\tau$ (s) | $\gg$ Landau-Khalatnikov time |
| Thickness of dielectric gap | $\lambda$ (nm) | 0.2 - 2 |

| Permittivity of the dielectric gap | $\varepsilon_d$ (dimensionless) | 1 - 10 |
| --- | --- | --- |
| Universal dielectric constant | $\varepsilon_0$ (F/m) | $8.85\times10^{-12}$ |
| Electron charge | $e$ (C) | $1.6\times10^{-19}$ |
| Ionization degree of the surface ions | $Z_i$ (dimensionless) | $Z_1 = +2,\ Z_2 = -2$ |
| Number of surface ions created per oxygen molecule | $n_i$ (dimensionless) | $n_1 = 2,\ n_2 = -2$ |
| Saturation area of the surface ions | $A_i$ (m²) | $A_1 = A_2 = 10^{-18}$ |
| Surface defect/ion formation energy | $\Delta G_i^{00}$ (eV) | 0.2 |

## APPENDIX D.

**Polarization curves, hysteresis loops and phase diagrams calculated using Gaussian Process model and 2-4 Landau phenomenological model**

Here, we calculate the objective function from order parameters and free energy to define and identify some phases mentioned in **Table I (main text).**

### D.1. Formulation of an Objective function

$$f_o = -1^m \left\{ \left(\frac{|p_1-p_2|}{\max(p_1,p_2)}\right) w_1 M + (m-1) w_2 M + \left(\frac{e_2-e_1}{\max(|e_1|,|e_2|)}\right) w_3 M \right\}, \quad (D.1)$$

$$m = \begin{cases} 0 & if\ A = 0; P = 0 \\ 1 & if\ A = 0; P = \pm x \\ 2 & if\ A = \pm x, P = 0 \\ 4 & if\ A = \pm x, P = \pm x \end{cases} \quad (D.2)$$

$$w_1 + w_2 + w_3 = 1 \quad (D.3)$$

Where $p_1, p_2$ are the absolute values of the maximum order parameters values (P, A) respectively from the list of minima points of free energy; $e_1, e_2$ are the respective minimum free energy values at $p_1, p_2$; $m$ is an indicator which differentiate phases with different range of objective values and is defined in 2; $w$ is the weighting parameter and $M$ is a large positive number which is used to tweak the objective function such that we penalize more as the objective $|f_o - 0|$ gets higher. This is done to efficiently use Gaussian Process model (**GPM**) for rapid exploration and prediction of phase diagram (provided later), as with negligible difference in the objectives of different phases, the GP can fail to detect the boundaries between phases. The last term of $f_o$ is to differentiate the FEI phase having deeper wells between the order parameters. That is to identify the location in the domain space, where non-zero P or A values provide global minima. Since for only FEI phase, where we have both the order parameters have values at minimal points, we ignore this term for other phases for simplicity. Thus,

$$\begin{cases} \left(\frac{e_2-e_1}{\max(|e_1|,|e_2|)}\right) & if\ A = \pm x, P = \pm x \\ 0 & else \end{cases} \quad (D.4)$$

### D.2. Summary of different phases in terms of objective function value

**PE:** minimal function value (negative) with $f_o = w_2 M$ where $w_1 < w_2$

**AFI:** function value (negative) increases than PE phase with $f_o = w_1 M$ where $w_1 < w_2$

**AFE:** function value increases than above two phases with $f_o = w_1 M + w_2 M$

**FEI:** maximum function value with $f_o = x_1 w_1 M + 3 w_2 M + x_2 w_3 M$ where $0 \leq x_1 \leq 1$ and $-1 \leq x_2 \leq 1$

In the FEI phase, objective value will be greater when P solutions will have deeper wells on minimum free energy (deeper minima) than A solutions than the vice versa.

**TABLE DI. Parameter values for Table I in the main text**

| Parameters | Designation and dimensionality | Value for a structure PbZrO$_3$ film / ionic charge / gap / tip |
|---|---|---|
| Electric voltage on the tip | $U$ (V) | 7 |
| Dielectric stiffness | $\alpha_{pT}$ ($\times 10^5$ C$^{-2}$·J·m/K) | 2.7969 |
| Curie temperature | $T_P$ (K) | 463.2 |
|  | $\alpha_{aT}$ ($\times 10^6$ C$^{-2}$·J·m/K) | 4.8789 |
| AFE temperature | $T_A$ (K) | 490 |
| Coefficient of LGD functional | $\beta_p$ ($\times 10^8$ J C$^{-6}$·m$^9$) | 3.825 |
| Coefficient of LGD functional | $\beta_a$ ($\times 10^9$ J C$^{-6}$·m$^9$) | 2.056 |
| Coefficient of LGD functional | $\chi$ ($\times 10^{10}$ J C$^{-6}$·m$^9$) | 1.7754 |
| Coefficient of LGD functional | $\gamma_p$ ($\times 10^9$ J C$^{-8}$·m$^{13}$) | 0 |
| Coefficient of LGD functional | $\gamma_a$ ($\times 10^{10}$ J C$^{-8}$·m$^{13}$) | 0 |
| Thickness of ferroelectric layer | $h$ (nm) | 50, 5 |
| Background permittivity of ferroelectric | $\varepsilon_{33}^b$ (dimensionless) | 7 |
| Thickness of dielectric gap | $\lambda$ (nm) | 2 |
| Permittivity of the dielectric gap | $\varepsilon_d$ (dimensionless) | 10 |
| Universal dielectric constant | $\varepsilon_0$ (F/m) | 8.85×10$^{-12}$ |
| Electron charge | $e$ (C) | 1.6×10$^{-19}$ |
| Ionization degree of the surface ions | $Z_i$ (dimensionless) | $Z_1 = +2$, $Z_2 = -2$ |
| Number of surface ions created per oxygen molecule | $n_i$ (dimensionless) | $n_1 = 2$, $n_2 = -2$ |
| Saturation area of the surface ions | $A_i$ (m$^2$) | $A_1 = A_2 = 10^{-18}$ |
| Surface defect/ion formation energy | $\Delta G_i^{00}$ (eV) | 0.2, 0.02 |
|  | $k_b$ | 1.38×10$^{-23}$ |

### D.3. Results

Below we have showed phase diagrams over the parameter space of temperature $T$ and relative partial oxygen pressure ρ. Different domain region has been explored with different values of thickness $h$ and Surface defect/ion formation energy $\Delta G_i^{00}$, considering the expression of free energy as per equation (6-7e) and to illustrate Table I for $\beta_{a,p} > 0$ and $\gamma_{a,p} = 0$. The parameter values considered to find the order parameters at the minimum points of the free energy equations is provided in Table C1. For each of the phase diagrams, we took two approaches – the first one is the exhaustive rectangular grid search where we have utilized 60 points per dimension, total of (60 x 60 =3600) grid locations. The second approach where we have used pure exploration to adaptively sample (Bayesian learning) only 10 % of the total grid points (360 points) and fit GPM to predict phases and phase transitions. For each figure, the first two image shows the mesh-grid of individual parameter space of $T$ and $\rho$. The third, fourth and fifth images are the phase diagram from exhaustive grid search, predicted phase diagram from GPM using 10% of grid utilization, the uncertainty of the GPM over the same parameter space respectively. X and Y axis for all the images in the figures are grid points and the color represents the parameter or function values such that black area has the lower values and white area has the higher values.

<u>Interpretation of **Figs. D1, D2, D3**.</u> In the true image (middle image), black region is the PE phase, green region is the AFE phase and yellow/red/white region is the FEI phase. In the FEI phase, yellow region is the area having deeper wells at A and as the objective value increases (with color changing from yellow to red to white), we have deeper wells at P. The domain does not have AFI phase.

<u>Interpretation of **Figs. D4**.</u> In the true image (middle image), black region is the PE phase, yellow region is the AFE phase and red/white region is the FEI phase. In the FEI phase, red region is the area having deeper wells at A and as the objective value increases (with color changing from red to white), we are progressing towards deeper wells at P. The domain does not have AFI phase.

In the GP predicted image (4th image of S1-S4), we can see similar distinctive region with different color coding and thus gave us the interpretation of individual phases. We can have better prediction of phases with more advanced acquisition function (exploration -exploitation) to sample where the objective function is higher.

<u>Comparison of phases diagrams in **Figs. D1-D4**.</u> We observed that as $\rho$ increase or decreases by the order of 10 for temperature $T < T_p$, we are approaching towards FEI phase where with further increase or decreases of $\rho$ by the order of 10, the deeper wells for order parameter A are shifted to deeper wells for order parameter P. Also, as the parameters $h$ and $\Delta G_i^{00}$ decreases, the FEI region shrinks and expands respectively for the same parameter space of $T, \rho$. Interestingly as per defined in Table I, we did not find the AFI region ($A = 0; P = \pm x$) for any of the figures.

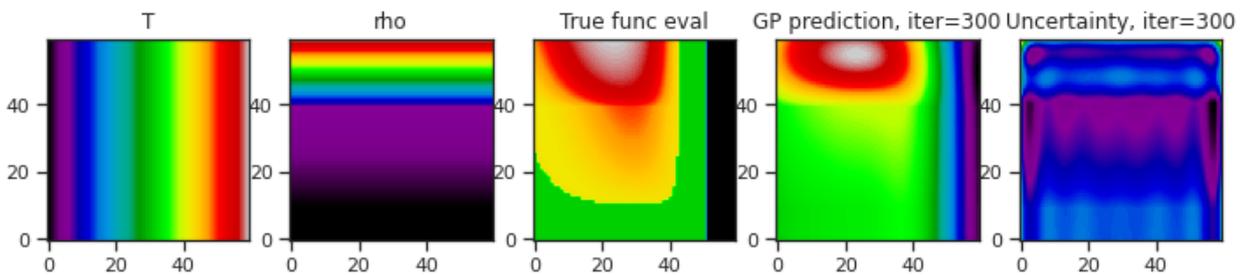

**FIGURE D1.** Phase Diagram: Parameter Space: $T = [300, 520]K$ and $\rho = [10^1, 10^6]$ with $E = 0$, thickness $h = 50$ nm, $\Delta G_i^{00} = 0.2$.

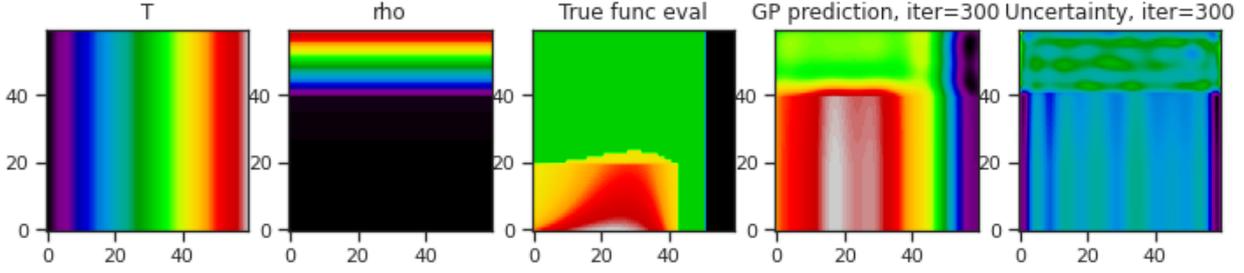

**FIGURE D2.** Phase diagram in the parameter space: $T = [300, 520]K$ and $\rho = [10^{-6}, 10^{-1}]$ with $E = 0$, thickness $h = 50$ nm, $\Delta G_i^{00} = 0.2$.

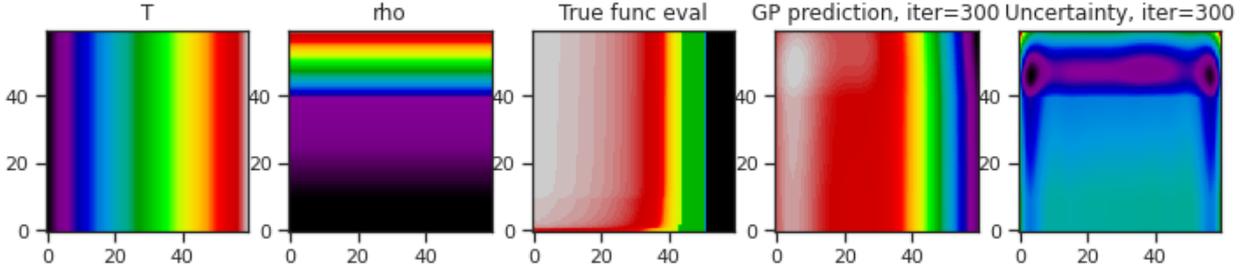

**FIGURE D3.** Phase diagram in the parameter space: $T = [300, 520]K$ and $\rho = [10^1, 10^6]$ with $E = 0$, thickness $h = 50$ nm, $\Delta G_i^{00} = 0.02$.

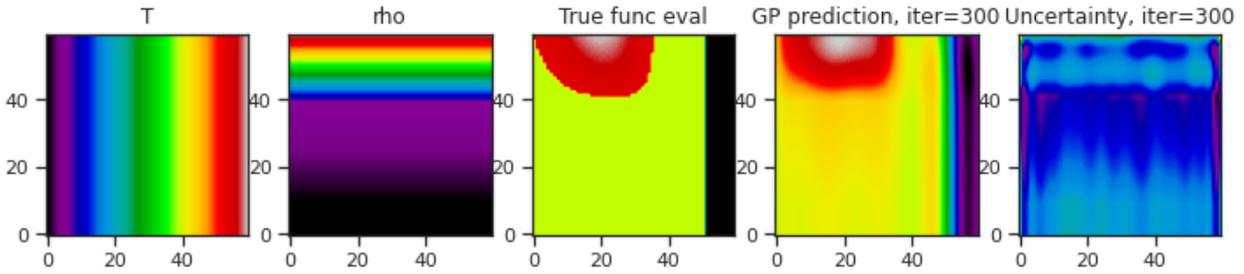

**FIGURE D4.** Phase diagram in the parameter space: $T = [300, 520]K$ and $\rho = [10^1, 10^6]$ with $E = 0$, thickness $h = 5$ nm, $\Delta G_i^{00} = 0.2$.

## APPENDIX E. Polarization curves, hysteresis loops and phase diagrams calculated using 2-4-6 Landau phenomenological model

The dependences of the anti-polarization $A$ (dashed and dotted blue loops) and polarization $P$ (dashed red and dark-red curves) on the static electric field $E$ are shown in **Fig. E1** for temperatures $T = (200 - 500)$ K (columns a-d) and relative oxygen pressures $\rho = 10^4, 1, 10^{-4}, 10^{-6}$ (from the top to the bottom rows), which values are listed for each column/row.

Polarization hysteresis loops $P(E)$ calculated for dimensionless frequencies $w = 0.3$ (black loops), 3 (red loops), 10 (magenta loops), 30 (blue loops), and 100 (green loops), and temperatures $T = (200 - 500)K$ and relative oxygen pressures $\rho = 10^4, 1, 10^{-4}, 10^{-6}$ are shown in **Fig. E2.**

A typical phase diagram of a thin AFE film in dependence on the temperature $T$ and partial oxygen pressure $\rho$ in shown in **Fig. E3**. There are an antiferroelectric (AFE) phase coexisting with a weak ferroelectric (FE) phase, a ferroelectric-like antiferroionic (AFI) phase, and an electret-like paraelectric (PE) phase. The light-blue region of AFE-FE coexisting phases is the widest and located at temperatures lower than 500 K (that is slightly higher than $T_A \cong 490$ K). The central part inside a slim white triangle, located in the AFE-FE region, tends to the diagram shown in **Fig. 2 (main text)**, and here an ultra-thin region of pure AFE phase exists around $\rho = 1$. The boundary between the AFE-FE phase region and two symmetrically located AFI regions is an almost straight inclined line. A ferroelectric-like FEI phase is located inside two relatively small (in comparison with AFE-FE region) light-green regions located at rather high (the top "hill", $\rho \geq 10^4$) or rather low (the bottom "hill", $\rho \leq 10^4$) relative oxygen pressures, and temperature range 320 K $\leq T \leq$ 500 K. The boundary between AFE-FE, AFI and electret-like PE phase (local inside a sand-colored region) is close to the vertical line $T \cong 500$ K. However, the boundary of PE phase has two small convexities from the straight line, which origin are unclear for us.

Three free energy maps and order parameter $E$-dependence (static hysteresis curves), calculated in each phase for the partial pressure $\rho = 10^{-6}$, are shown in the top and bottom insets, respectively. The top insets illustrate the typical free energy relief is AFE-FE, FE-like FEI and electret-like PE phase, respectively. The bottom insets are corresponding hysteresis loops. The description of the insets is the same as in **Figs. 3** and **4 (main text)**, respectively.

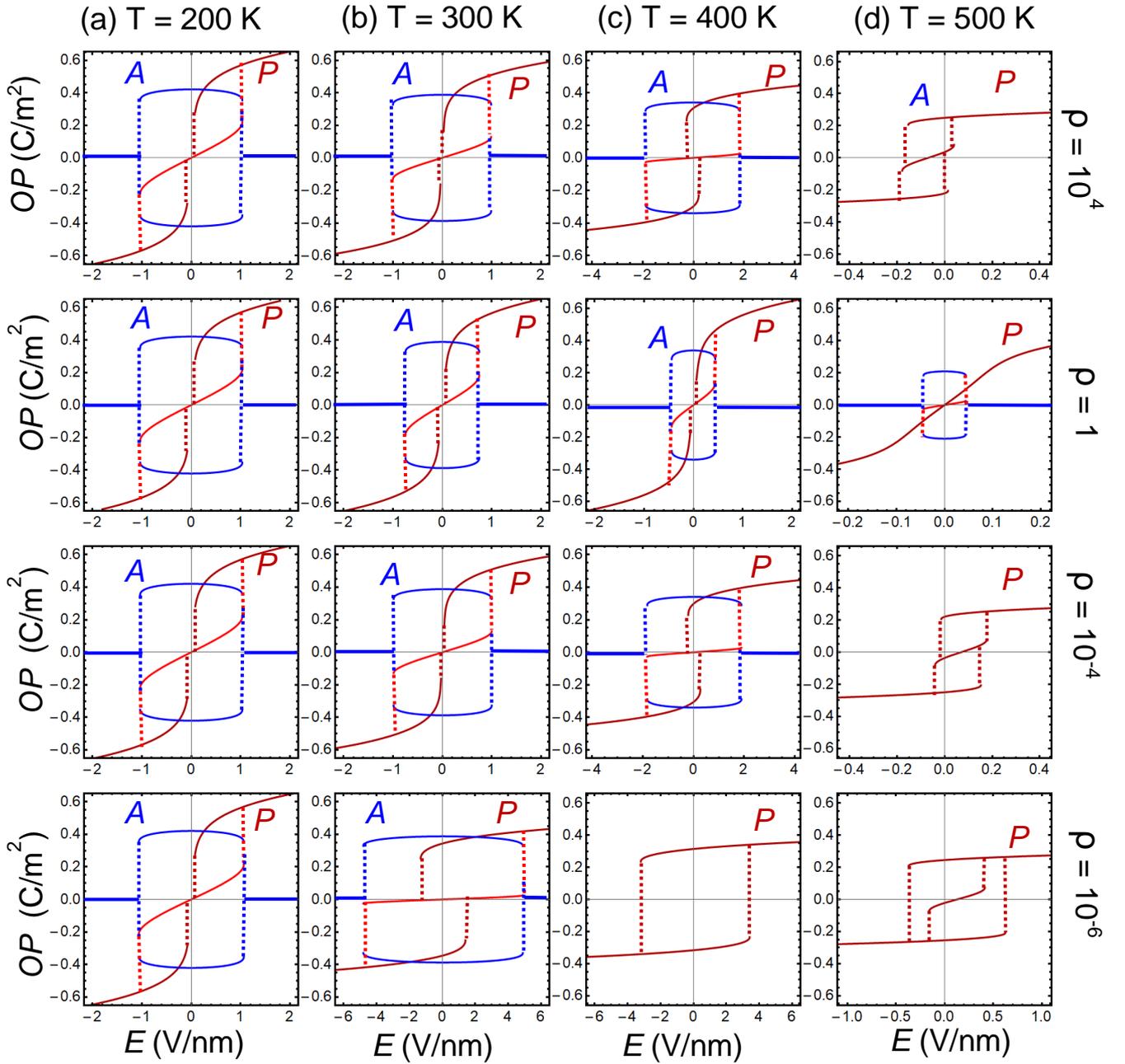

**FIGURE E1.** Static dependences of the order parameters (OP) – antipolar parameter $A$ (solid blue curves) and polarization $P$ (solid red and dark-red curves) on applied electric field $E$ calculated for temperature $T =$ 200, 300, 400 and 500 K [columns **(a)**, **(b)**, **(c)** and **(d)**] and relative partial oxygen pressures $\rho = 10^4, 1, 10^{-4}, 10^{-6}$, which values are listed for each column/row. Dotted vertical lines show the thermodynamical transitions between different polar and anti-polar states. An external electric field is $E = \frac{U}{h}$. Other parameters are the same as in **Fig. 3.**

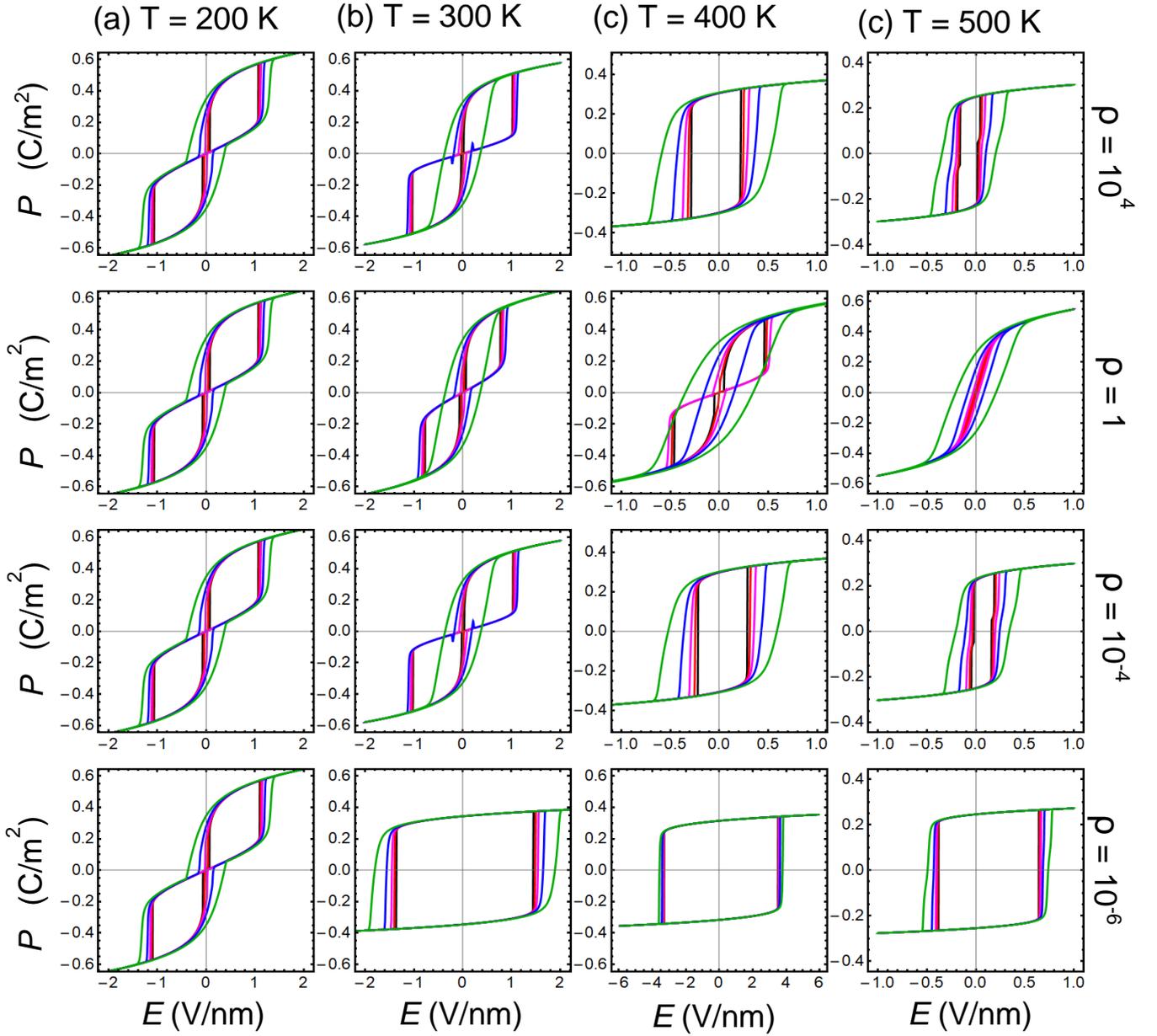

**FIGURE E2.** Polarization hysteresis $P(E)$ calculated for dimensionless frequencies $w = 0.3$ (black loops), 3 (red loops), 10 (magenta loops), 30 (blue loops), and 100 (green loops), and temperatures $T = 200$ K, 300 K, 400 K and 500 K [columns **(a), (b), (c)** and **(d)**, respectively] and relative partial oxygen pressures $\rho = 10^4, 1, 10^{-4}, 10^{-6}$, which values are listed for each column/row. External electric field is $E = \frac{U_0}{h}\sin(\omega t)$, and $w = \frac{\omega \Gamma_P}{2|\alpha_p|}$. Other parameters are the same as in Fig. 3.

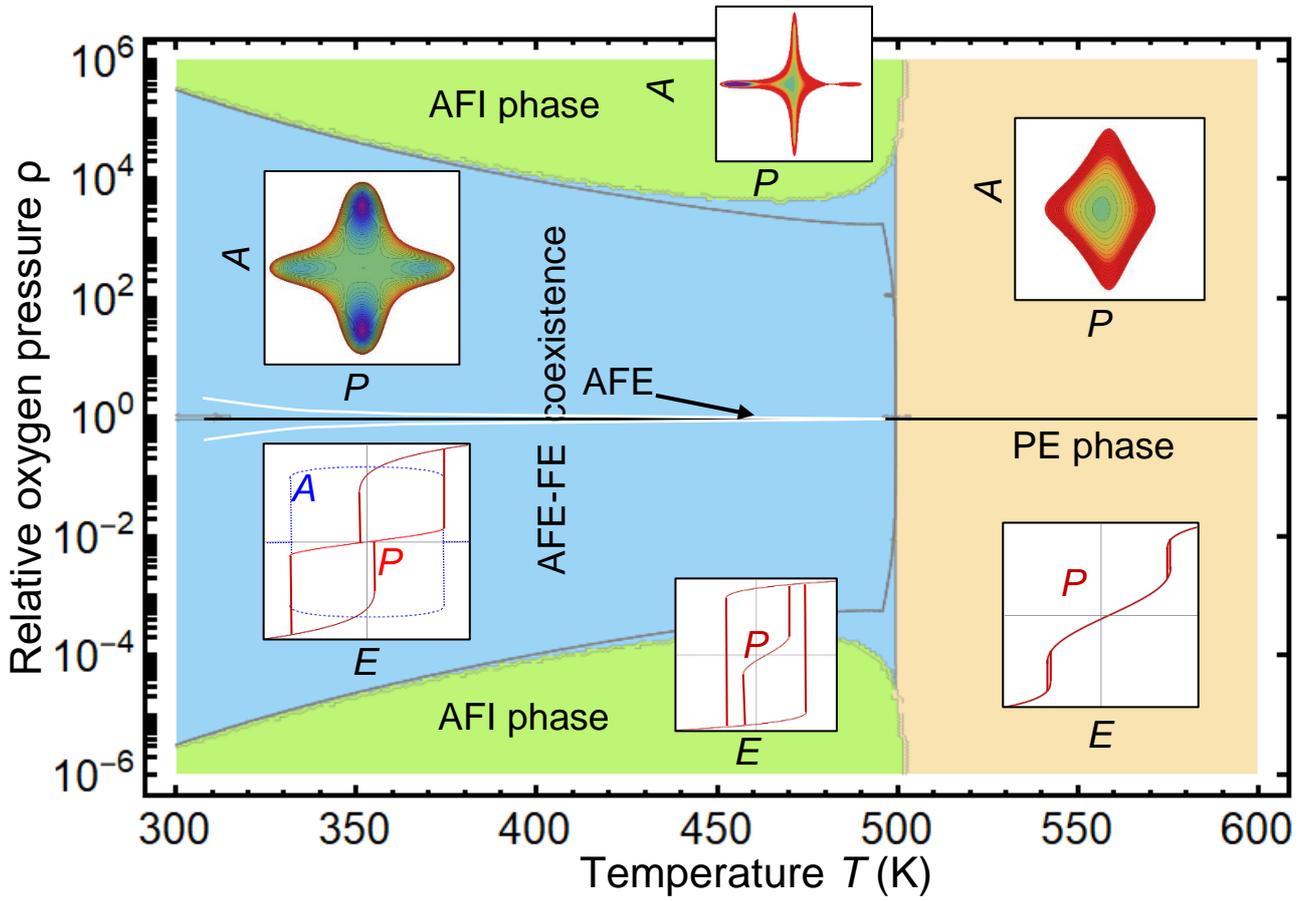

**FIGURE E3.** A phase diagram of a thin AFE film in dependence on the temperature $T$ and relative partial oxygen pressure $\rho$. There are an antiferroelectric (AFE) phase coexisting with a weak ferroelectric (FE) phase, a ferroelectric-like antiferroionic (AFI) phase, and an electret-like paraelectric (PE) phase. Free energy maps at $E = 0$ (upper insets) and order parameter dependence on the static $E$-field (bottom insets), are calculated in each phase for the partial oxygen pressure $\rho = 10^{-6}$. The description of the insets is the same as in **Figs. 3** and **4**, respectively. Other parameters are the same as in **Fig. 3.**

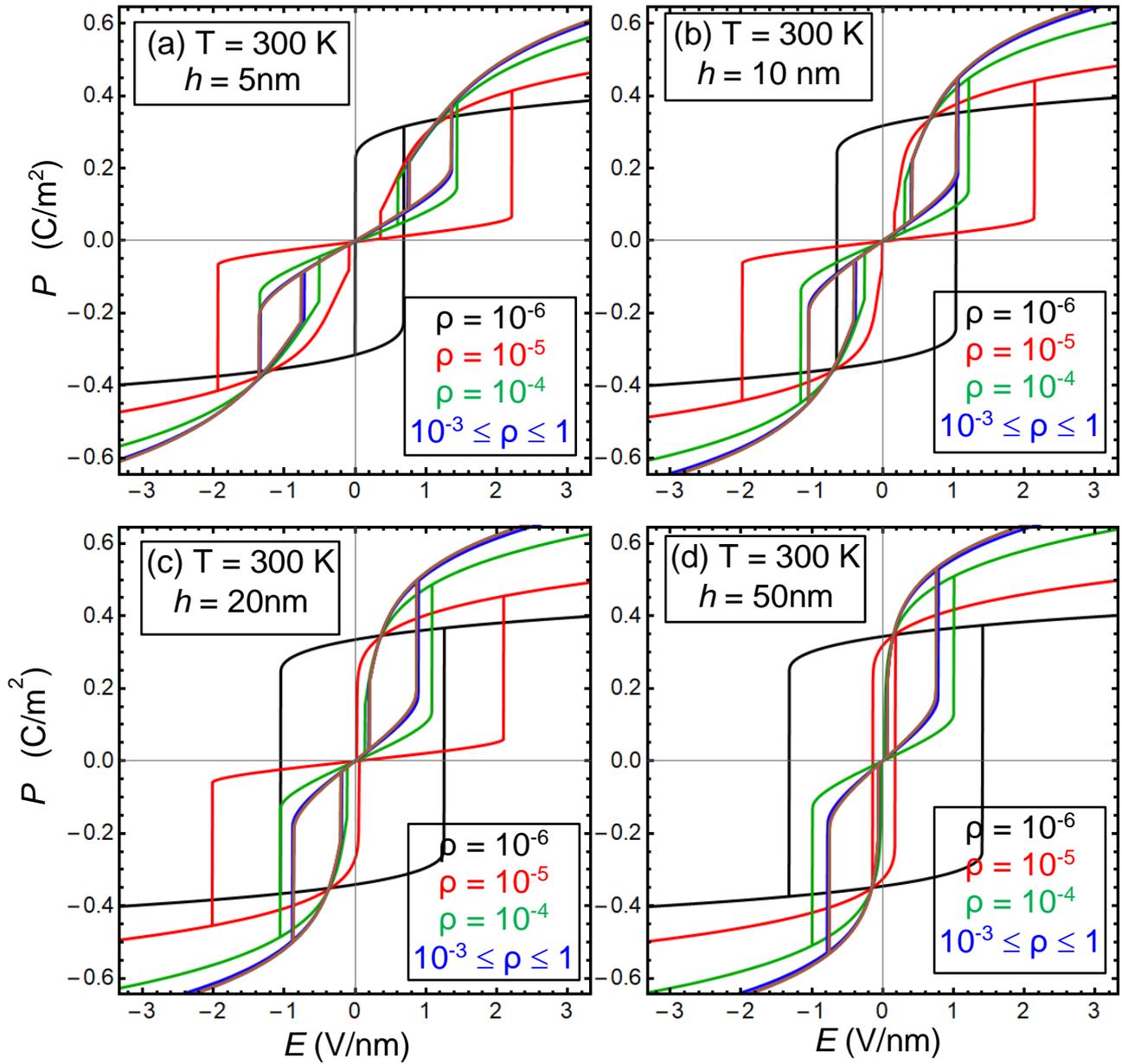

**FIGURE E4.** Polarization hysteresis $P(E)$ calculated for low dimensionless frequencies $w = 0.3$, room temperature $T = 300$ K, and relative partial oxygen pressure $\rho = 10^{-6}$, (black loops), $10^{-5}$ (red loops), $10^{-4}$ (green loops), and $10^{-3} \leq \rho \leq 1$, (blue and brown loops). The film thickness $h = 5$ nm **(a)**, 10 nm **(b)**, 20 nm **(c)** and 50 nm **(d)**. External electric field is $E = \frac{U_0}{h}\sin(\omega t)$, and $w = \frac{\omega \Gamma_P}{2|\alpha_p|}$. Other parameters are the same as in **Fig. 3.**

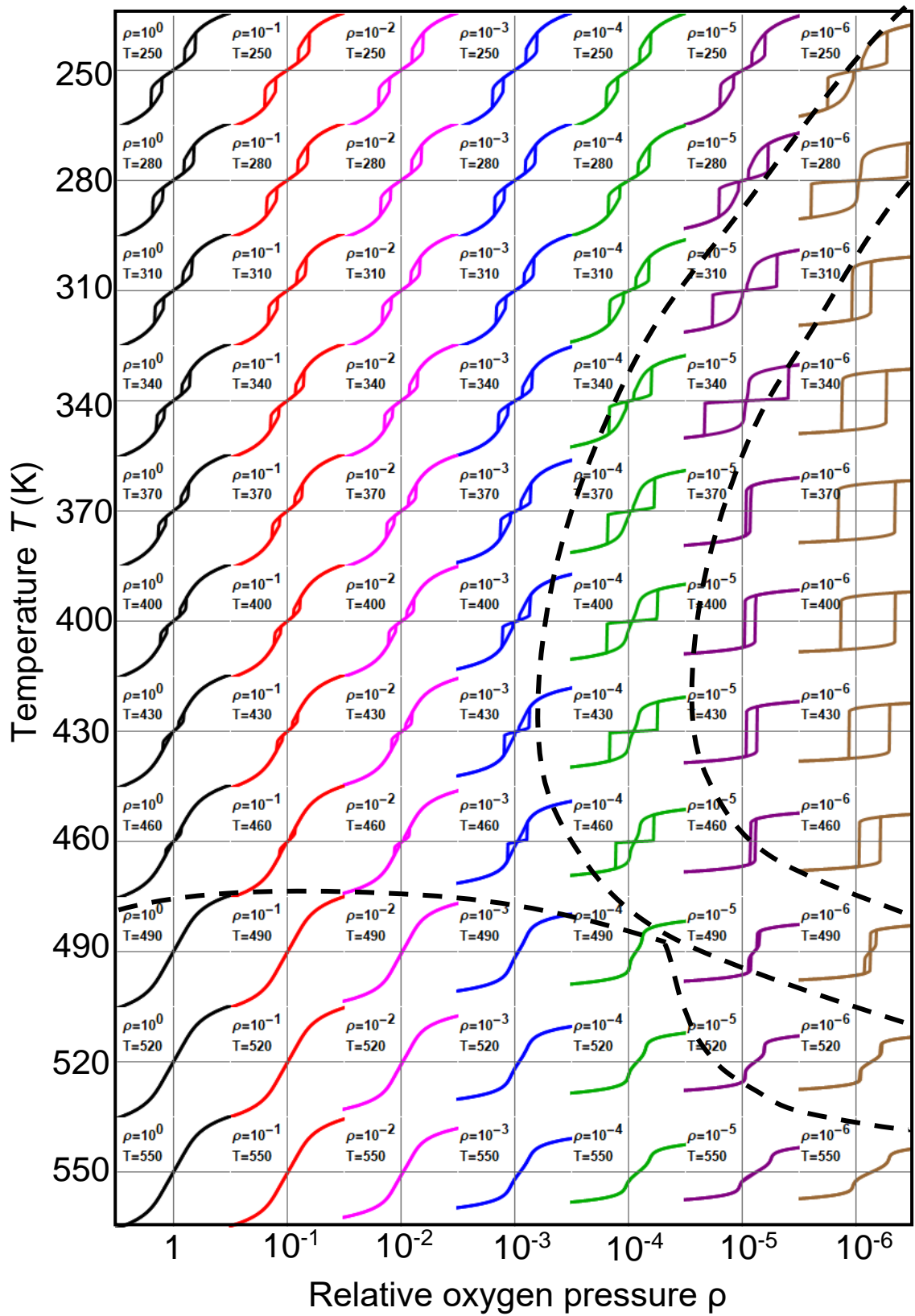

**FIGURE E5.** A schematic diagram of the loop shape in dependence on the relative pressure $\rho$ and temperature $T$. The film thickness $h = 5$ nm. Other parameters are the same as in **Fig. 3**. Dashed curves separate different type of loops explained in **Fig. 8** in the main text.

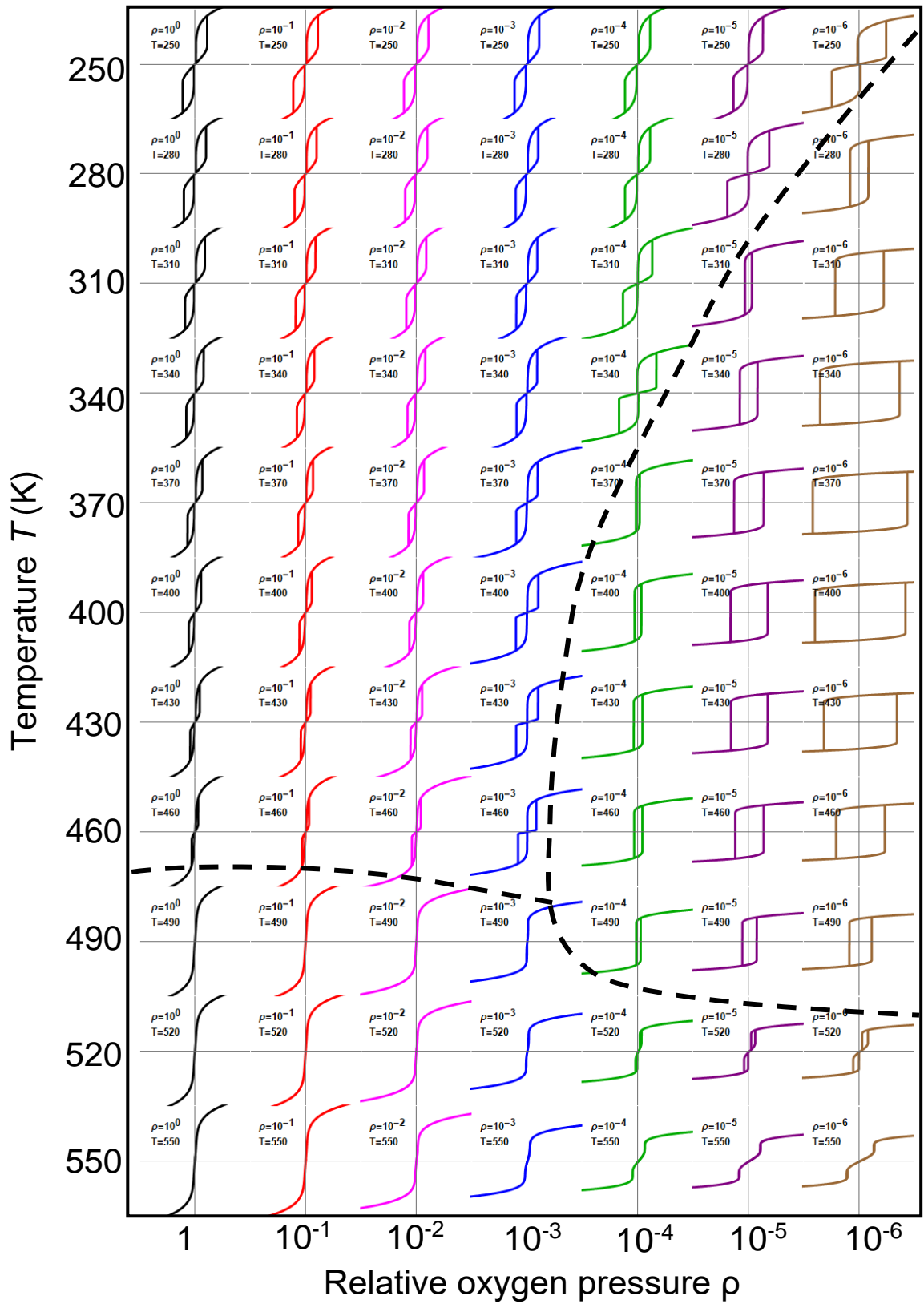

**FIGURE E6.** A schematic diagram of the loop shape in dependence on the relative pressure $\rho$ and temperature $T$. The film thickness $h = 50$ nm. Other parameters are the same as in **Fig. 3.** Dashed curves separate different type of loops explained in **Fig. 8** in the main text.

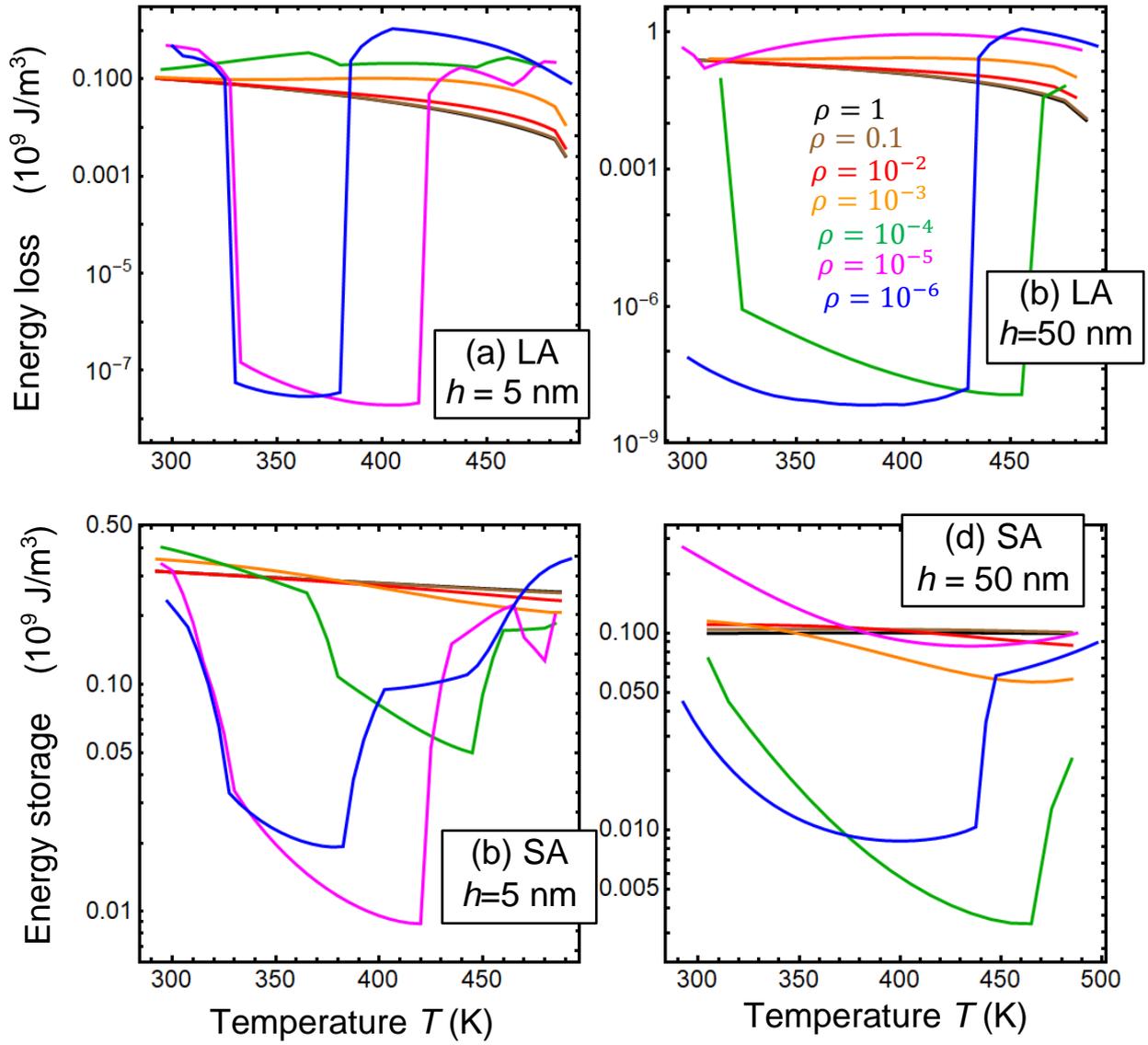

**FIGURE E7.** Temperature dependence of the energy loss (**LA**, plots **a,b**), and the stored energy (**SA**, plots **c,d**) for different values of relative pressure $\rho = 10^{-6}, 10^{-5}, 10^{-4}, 10^{-3}, 10^{-2}, 0.1$ and $1$ (blue, magenta, green, orange, red, brown, and black curves, respectively). The film thickness h=5 nm (**a, c**), and 50 nm (**b, d**). The dimensionless frequency $w = 10^{-3}$; other parameters are the same as in **Fig. 3**.

# REFERENCES

[1] M.J. Haun, Z.Q. Zhuang, E. Furman, S.J. Jang, and L.E. Cross, Thermodynamic theory of the lead zirconate-titanate solid solution system, Part III: Curie constant and sixth-order polarization interaction dielectric stiffness coefficients, Ferroelectrics, **99**, 45 (1989).